\documentclass[a4paper,11pt]{article}
\pdfoutput=1
\usepackage{jheppub}
\usepackage{xspace}

\usepackage[T1]{fontenc}

\newcommand{\bbbar}{\ensuremath{b\bar{b}}}

\newcommand{\Candidate}{{\tt Candidate}}
\newcommand{\TObjArray}{{\tt TObjArray}}
\newcommand{\TClonesArray}{{\tt TClonesArray}}
\newcommand{\TRef}{{\tt TRef}}
\newcommand{\TRefArray}{{\tt TRefArray}}

\newcommand{\etc}{\mbox{etc.}}

\newcommand{\PYTHIA}{\textsc{Pythia6}\xspace}
\newcommand{\PYTHIAEIGHT}{\textsc{pythia8}\xspace}
\newcommand{\MG}{\textsc{MadGraph5}\xspace}
\newcommand{\MGZ}{\textsc{MadGraph}\xspace}
\newcommand{\ROOT}{\textsc{Root}\xspace}
\newcommand{\LHCO}{\textsc{LHCO}\xspace}

\newcommand{\DELPHES}{\textsc{Delphes}\xspace}
\newcommand{\DELPHESTZE}{\textsc{Delphes 3.0.11}\xspace}

\newcommand{\FASTJET}{\textsc{FastJet}\xspace}
\newcommand{\GEANT}{\textsc{Geant}\xspace}

\title{\boldmath DELPHES 3 \\ A modular framework for fast simulation of a generic collider experiment}

\collaborationImg{\includegraphics[width=0.20\textwidth]{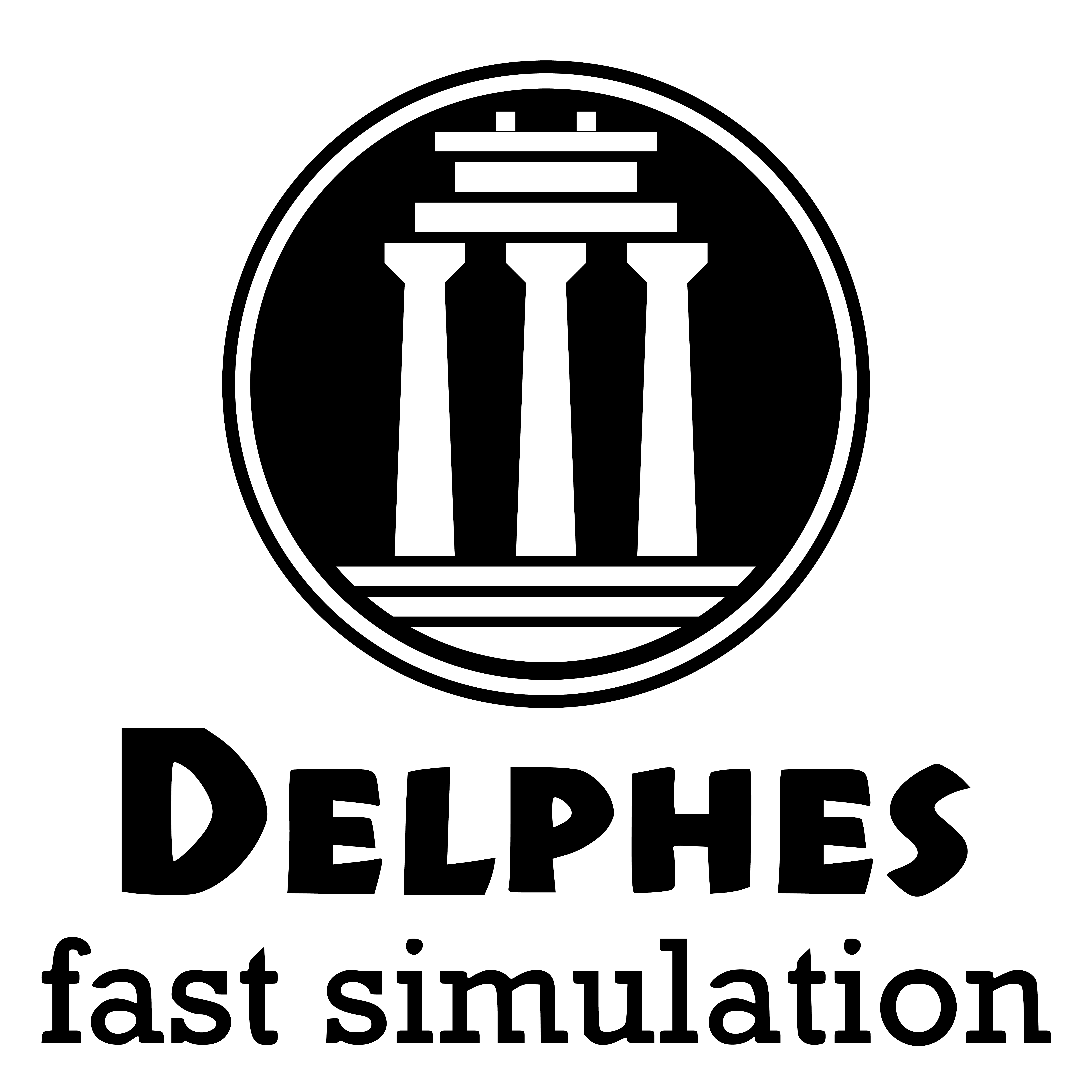}}

\author{J. de Favereau}
\author{, C. Delaere}
\author{, P. Demin}
\author{, A. Giammanco}
\author{, V. Lema\^itre}
\author{, A. Mertens}
\author{and M. Selvaggi}

\affiliation{Centre for Cosmology, Particle Physics and Phenomenology (CP3), \\
Universit\'e Catholique de Louvain, \\
Chemin du Cyclotron 2, B-1348 Louvain-la-Neuve, Belgium}

\emailAdd{jerome.defavereau@uclouvain.be}
\emailAdd{christophe.delaere@uclouvain.be}
\emailAdd{pavel.demin@uclouvain.be}
\emailAdd{andrea.giammanco@uclouvain.be}
\emailAdd{vincent.lemaitre@uclouvain.be}
\emailAdd{alexandre.mertens@uclouvain.be}
\emailAdd{michele.selvaggi@uclouvain.be}

\abstract{The version 3.0 of the \DELPHES fast-simulation is presented. The goal of \DELPHES is to allow the simulation of a multipurpose detector for phenomenological studies. The simulation includes a track propagation system embedded in a magnetic field, electromagnetic and hadron calorimeters, and a muon identification system. Physics objects that can be used for data analysis are then reconstructed from the simulated detector response. These include tracks and calorimeter deposits and high level objects such as isolated electrons, jets, taus, and missing energy. The new modular approach allows for greater flexibility in the design of the simulation and reconstruction sequence. New features such as the particle-flow reconstruction approach, crucial in the first years of the LHC, and pile-up simulation and mitigation, which is needed for the simulation of the LHC detectors in the near future, have also been implemented. The \DELPHES framework is not meant to be used for advanced detector studies, for which more accurate tools are needed.  Although some aspects of \DELPHES are hadron collider specific, it is flexible enough to be adapted to the needs of electron-positron collider experiments. }

\begin{document} 
\maketitle
\flushbottom

\section{Introduction}\label{sec:intro}

High energy particle collisions can produce a large variety of final states. Highly sophisticated detectors are designed in order to detect and precisely measure particles originating from such collisions. Experimental collaborations often rely on Monte-Carlo event generation for designing and optimizing specific analysis strategies. Whenever such studies require a high level of accuracy, the interactions of long-lived particles with the detector matter content are fully simulated with the \GEANT package~\cite{bib:geant4}, electronic response is emulated by dedicated routines, and final observables are reconstructed by means of complex algorithms. To face the limited computing resources and still allow the use of large samples (for example when scanning parameter spaces), the LHC collaborations have developed fast-simulation techniques~\cite{bib:atlfast1,bib:atlfast2,bib:cmsfast1,bib:cmsfast2,bib:cmsfast3} which are two to three orders of magnitude faster than the fully GEANT based simulations.

These procedures require expertise and the deployment of large scale computing resources that can be handled only by large collaborations. For most phenomenological studies, such a level of complexity is not needed and a simplified approach based on the parameterization of the detector response is in general good enough. In 2009, the \DELPHES framework~\cite{bib:delphes} was designed to achieve such goal. The \DELPHES framwork takes as input the most common event generator output and performs a fast and realistic simulation of a general purpose collider detector. To do so, long-lived particles emerging from the hard scattering are propagated to the calorimeters within a uniform magnetic field parallel to the beam direction. The particle energies are computed by smearing the initial long-lived visible particles momenta according to the detector resolution. As a result, jets, missing energy, isolated electrons, muons and photons, and taus can be reconstructed.

With respect to its previous incarnation~\cite{bib:delphes}, the present version of
\DELPHES includes an attempt to roughly emulate the particle-flow reconstruction philosophy used in ALEPH~\cite{bib:eflow} 
and CMS~\cite{bib:pflow}, based on the optimally-combined use of the information from all the subdetectors to
reconstruct and identify all particles indvidually. While the aim is not
to re-implement the particle-flow algorithm in all its complexity (for
example, electrons, muons, and photons, are assumed to be perfectly
identified and have no fake rate in \DELPHES), the simplified approach
adopted here is particularly suitable for the treatment of pile-up, as well
as for the emulation of $\rm{b}$ and $\rm{\tau}$ tagging, and is able to reproduce the
jet and missing energy resolutions observed in CMS with their complete
reconstruction. From a technical perspective, the code structure is now fully modular, providing greater flexibility to the user and allowing the integration of \DELPHES routines in other projects. 

The modeling of the detector response, as well as the reconstruction and validation of the physical observables are described in Sections~\ref{sec:detsim},~\ref{sec:objrec},~\ref{sec:hlobj} and~\ref{sec:validation}. A couple of illustrative use cases of \DELPHES in the context of LHC studies are presented in Section~\ref{sec:usecase}. Although some aspects presented in the following are hadron collider specific, such as the use of transverse variables, \DELPHES is flexible enough to be adapted to the needs of electron-positron collider experiments (see Sections~\ref{sec:isolation} and~\ref{sec:metht}). 

\section{Simulation of the detector response}\label{sec:detsim}

The \DELPHES framework simulates the response of a detector composed of an inner tracker, electromagnetic and hadron calorimeters, and a muon system. 
All are organized concentrically with a cylindrical symmetry around the beam axis. 
The user may specify the detector active volume, the calorimeter segmentation and the strength of the uniform magnetic field. 
Each sub-detector has a specific response, as described in the following. 

\subsection{Particle Propagation}\label{sec:partprop}

The first step carried by \DELPHES is the propagation of long-lived particles within a uniform axial magnetic field parallel to the beam direction. The magnetic field is assumed to be localized in the inner tracker volume. If the particle is neutral, its trajectory is a straight line from the production point to a calorimeter cell. If it is charged, it follows a helicoidal trajectory until it reaches the calorimeters. Particles that originate from a point outside the tracker volume are ignored. 

Charged particles have a user-defined probability to be reconstructed as tracks in the central tracking volume. 
A perfect angular resolution on tracks is assumed, therefore only a smearing on the norm of the transverse momentum vector is applied at the stage of particle propagation. 
This hypothesis is valid for most of the past, present and future particle detectors.
As for the tracking efficiency, energy and momentum resolutions can be specified by the user and depend on the particle type, transverse momentum and pseudo-rapidity.

\subsection{Calorimeters}\label{sec:calo}

After their propagation in the magnetic field, long-lived particles reach the calorimeters. The electromagnetic calorimeter, ECAL, is responsible for measuring the energy of electrons and photons, while the hadron calorimeter, HCAL, measures the energy of long-lived charged and neutral hadrons. 

In \DELPHES, the calorimeters have a finite segmentation in pseudo-rapidity and azimuthal angle ($\eta$,$\phi$). The size of the elementary cells can be defined in the configuration file. For simplicity the segmentation is uniform in $\phi$ and for computational reasons we assume the same granularity for ECAL and HCAL. The coordinate of the resulting calorimeter energy deposit, the tower, is computed as the geometrical centre of the cell. 

Long-lived particles reaching the calorimeters deposit a fixed fraction of their energy in the corresponding ECAL ($f_{ECAL}$) and HCAL ($f_{HCAL}$) cells. Since ECAL and HCAL are perfectly overlaid, each particle reaches one ECAL and one HCAL cell. The resulting ECAL and HCAL cells are grouped in a calorimeter tower. By default, in \DELPHES, electrons and photons leave all their energy in ECAL ($f_{ECAL}=1$). Although in a real detector stable hadrons deposit a significant fraction of their energy in ECAL, in \DELPHES we assume that all their energy is deposited in HCAL ($f_{HCAL}=1$). Kaons and $\rm{\Lambda}$ have a finite lifetime but are considered stable by most event generators. In \DELPHES, rather than decaying these particles we assume that they share their energy deposit between ECAL and HCAL. The values $f_{ECAL}=0.3$ and $f_{ECAL}=0.7$ have been chosen according to the dominant decay products of such particles~\cite{bib:pdg}. Muons, neutrinos and neutralinos, do not deposit energy in the calorimeters. 
The user has the freedom to change the default setup, and define for each long-lived particle more accurate values for $f_{ECAL}$ and $f_{HCAL}$.

The resolutions of ECAL and HCAL are independently parameterized as a function of the particle energy and the pseudo-rapidity:
\begin{equation}
\left(\frac{\sigma}{E}\right)^2 = \left(\frac{S(\eta)}{\sqrt{E}}\right)^2 
                                                        + \left(\frac{N(\eta)}{E}\right)^2
                                                        + C(\eta)^2\,,
\label{eq:calores}
\end{equation}
where $S$, $N$ and $C$ are respectively the \textit{stochastic}, \textit{noise} and \textit{constant} terms. The electromagnetic and hadronic energy deposits are independently smeared by a log-normal distribution. The final tower energy is then computed as: 
\begin{equation}
E_{Tower} =  \sum_{particles}\text{ln}\mathcal{N}\left(f_{ECAL} \cdot E,\sigma_{ECAL}(E,\eta)\right) +\text{ln}\mathcal{N}\left(f_{HCAL} \cdot E,\sigma_{HCAL}(E,\eta)\right) .
\label{eq:etow}
\end{equation}
The energy of each particle is concentrated in one single tower and the sum runs over all particles that reach the given tower. $\text{ln}\mathcal{N}(m,s)$ is the log-normal distribution with mean $m$ and variance $s$. The parameters $\sigma_{ECAL}$ and $\sigma_{HCAL}$ are respectively the ECAL and HCAL resolutions, defined in equation~(\ref{eq:calores}). A calorimeter tower is also characterized by its position in the ($\eta$,$\phi$) plane, given by the geometrical centre of the corresponding cell. In order to avoid having to deal with discrete tower positions, an additional uniform smearing of the position over the cell range is performed.

Calorimeter towers are, along with tracks, crucial ingredients for reconstructing isolated electrons and photons, as well as high-level objects such as jets and missing transverse energy. 

\subsection{Particle-flow Reconstruction}\label{sec:pfrec}

The philosophy of the particle-flow approach is to use a maximum amount of information provided by the various sub-detectors for reconstructing the event. This modus operandi has been adopted by some experimental collaborations (see for example~\cite{bib:eflow,bib:pflow}) but depends on the specificity of the experimental device. In \DELPHES, we have opted for a simplified approach based on the tracking system and the calorimeters for implementing the particle-flow event reconstruction. 

If the momentum resolution of the tracking system is better than the energy resolution of calorimeters, it can be convenient to use the tracking information within the tracker acceptance for estimating the charged particles momenta. In real experiments, the tracker resolution is better than the calorimeter resolution only up to some energy threshold. In the context of particle-flow reconstruction, we assume it is always convenient to estimate charged particle momenta via the tracker.

The particle-flow algorithm produces two collections of 4-vectors --- \emph{particle-flow tracks} and \emph{particle-flow towers} --- that serve later as input for reconstructing high resolution jets and missing transverse energy. 
For each calorimeter tower, the algorithm counts: 

\begin{itemize}
\item $E_{ECAL}$ and $E_{HCAL}$, the total energy deposited in ECAL and HCAL respectively. 
\item $E_{ECAL,trk}$ and $E_{HCAL,trk}$, the total energy deposited respectively in ECAL and HCAL originating from charged particles for which the track has been reconstructed. The charged components $E_{ECAL,trk}$ and $E_{HCAL,trk}$ can be 
asserted if one assumes perfect charged particle identification for reconstructed tracks. 
\end{itemize}

We then define:

\begin{equation}
\Delta_{ECAL} =E_{ECAL} - E_{ECAL,trk}\,,
\qquad
\Delta_{HCAL} =E_{HCAL} - E_{HCAL,trk}\,,
\label{eq:efexcess}
\end{equation}
\begin{equation}
E^{eflow}_{Tower} =  \max(0,\Delta_{ECAL}) + \max(0,\Delta_{HCAL})
\label{eq:eftower}
\end{equation}
\\

The particle-flow proceeds then as follows: 

\begin{itemize}
\item each reconstructed track results in an \emph{particle-flow track}
\item if $E^{eflow}_{Tower}>0$, an \emph{particle-flow tower} is created with energy $E^{eflow}_{Tower}$.
\end{itemize}

To better illustrate the particle-flow algorithm in \DELPHES here are a few simple examples:

\begin{itemize}
\item a single charged pion is reconstructed as a track with energy $E_{HCAL,trk}$ and deposits some energy $E_{HCAL}$ in the HCAL. If $E_{HCAL} \leq E_{HCAL,trk}$ only a particle-flow track with energy $E_{HCAL,trk}$ is produced. If $E_{HCAL} >  E_{HCAL,trk}$, a particle-flow track with energy $E_{HCAL,trk}$ and a particle-flow tower with energy $E_{HCAL}$ are produced. 
\item a single photon deposits its energy $E_{ECAL}$ in an ECAL cell. No tracks pointing to the cell is reconstructed. A particle-flow tower is created with energy $E_{ECAL}$. 
\item a photon and a charged pion reach the same calorimeter tower, the former deposits some energy $E_{ECAL}$ in ECAL and the latter $E_{HCAL}$ in HCAL. Furthermore, the charged pion is reconstructed as a track, with an energy $E_{HCAL,trk}$. If $E_{HCAL} \leq E_{HCAL,trk}$, a particle-flow track with energy $E_{HCAL,trk}$ and a particle-flow tower with energy $E_{ECAL}$ are created. If $E_{HCAL} > E_{HCAL,trk}$, a particle-flow track with energy $E_{HCAL,trk}$ and a particle-flow tower with energy $E_{ECAL} + E_{HCAL} - E_{HCAL,trk}$ are created. 
\end{itemize}

Defined that way, the particle-flow tracks contain charged particles estimated with a good resolution, while the particle-flow towers contain in general a combination of neutral particles, charged particles with no corresponding reconstructed track and additional excess deposits induced by the positive smearing of the calorimeters, and are characterized by a lower resolution. As shown in sections~\ref{sec:jets} and~\ref{sec:pus}, besides producing high-resolution inputs for jets and missing transverse energy, the particle-flow approach can be rather useful for addressing pile-up subtraction. While very simple when compared to what is actually required in real experiments, the algorithm described above is shown to reproduce well the performance achieved at LHC later in section~\ref{sec:validation}.

\section{Object Reconstruction}\label{sec:objrec}

In \DELPHES, the object reconstruction and identification is based on a series of approximations to sensibly speed up the procedure while keeping good accuracy. 

\subsection{Charged leptons and photons}\label{sec:chleppho}

\subsubsection{Charged leptons}\label{sec:chlep}

\paragraph{Taus} Hereafter, since $\rm{\tau}$ leptons decay before being detected, we refer by charged leptons solely to electrons ($\rm{e^\pm}$) and muons ($\rm{\mu^\pm}$). 
The reconstruction of hadronically decaying $\rm{\tau}$'s is addressed in section~\ref{sec:btau}.

\paragraph{Muons} In \DELPHES, a muon originating from the interaction, has some probability of being reconstructed, according to the user defined efficiency parameterization. This probability vanishes outside the tracker acceptance, and for muon momenta below some threshold to reject looping particles. The final muon momentum is obtained by a Gaussian smearing of the initial 4-momentum vector. The resolution is parameterized as a function of $p_T$ and $\eta$ by the user. 

\paragraph{Electrons} The full electron reconstruction usually involves combining information from the tracking system together with the electromagnetic calorimeter. In \DELPHES, we circumvent these reconstruction complexities by parameterizing the combined reconstruction efficiency as a function of the energy  and pseudorapidity. As for muons, the electron reconstruction efficiency vanishes outside the tracker acceptance and below some energy threshold. For the electron energy resolution, we use a combination of the ECAL and tracker resolution. At low energy, the tracker resolution dominates, while at high energy, the ECAL energy resolution dominates. 

\subsubsection{Photons}\label{sec:photons}

The reconstruction of photons relies solely on the ECAL. Photon conversions into electron-positron pairs are neglected. The final photon energy is obtained by applying the ECAL resolution function presented in section~\ref{sec:calo}. True photons and electrons with no reconstructed track that reach ECAL are reconstructed as photons in \DELPHES. \\ 

It is important to note that the fake rate for electrons, muons and photons is not simulated in the present \DELPHES version, as this feature goes beyond the scope of phenomenological applications. However, thanks to the modular structure of the framework, it is possible to implement in future versions a module that produces fake particles. The actual implementation and parametrization of the fake rate would nevertheless require a detailed input from the experimental collaborations.  

\subsubsection{Isolation}\label{sec:isolation}

An electron, muon or photon is isolated if the activity in its vicinity is small enough.
An isolated object has a small probability to originate from a jet. 
Several possible definitions exist for an isolation variable, depending on the particular level of signal-to-background rejection that the analyzer desires to achieve. 
In \DELPHES, we have opted for a simple one, well suited to hadron collider experiments. An alternative definition, more suitable to $\rm{e^+ e^-}$ experiments, based on spherical variables, although not yet implemented in \DELPHES, can be easily derived from the present one.  Moreover, the modularity of the framework allows the user other definitions, more suitable to different experiments or analysis requirements, or simply to not apply any isolation criteria on the final objects. 

For each reconstructed electron, muon, or photon ($\rm{P=e,\mu,\gamma}$), we define the isolation variable $I$ as:
\begin{equation}
I(P) = \frac{\displaystyle \sum_{i\neq P}^{\Delta R < R,\,\,p_T(i)>p_T^{min}} p_T(i)}{p_T(P)},
\label{eq:isolation}
\end{equation}
where the denominator is the transverse momentum of the particle of interest $\rm{P}$. The numerator is the sum of transverse momenta above $p_T^{min}$ of all particles that lie within a cone of radius $R$ around the particle $\rm{P}$, except $\rm{P}$. The input particle collection  entering the sum can be freely specified by the user. Particle-flow objects, or simply tracks and calorimeter towers are common choices for the input collection entering the isolation variable $I(P)$ calculation. Typically values of $I\approx 0$ indicate that the particle is \emph{isolated}. In \DELPHES, $\rm{P}$ is said to be \emph{isolated} if $I(P)<I_{min}$. The user can specify via the configuration file the three isolation parameters $p_T^{min}$, $R$ and $I_{min}$. The default values are $p_T^{min}=0.1$ GeV, $R=0.5$, $I_{min}=0.1$.

\subsection{Jets}\label{sec:jets}

\subsubsection{Jet reconstruction}\label{sec:jetrec}

In a hadron collider experiment, final states are often dominated by jets. 
An accurate jet reconstruction is therefore crucial. 
A naive approach would consist in parameterizing the jet response from the generated parton to the reconstructed jet. Although very fast, this approach would require constant input for tuning from real experiments and would have to be repeated for each variation of the jet reconstruction algorithms.
Moreover, such a parameterization would suffer from being process dependent and would not easily cope with extra radiation, hadronization and pile-up effects. 

Thanks to the modularity of the version 3 of \DELPHES, it is possible to produce jets starting from different input collections:
\begin{itemize}
\item \emph{Generated Jets} are clustered from generator level long-lived particles obtained after parton-shower and hadronization. No detector simulation nor reconstruction is taken into account.
\item \emph{Calorimeter Jets} use calorimeter towers defined in section~\ref{sec:calo} as input. 
\item \emph{Particle-flow Jets} are the result of clustering the particle-flow tracks and particle-flow towers defined in section~\ref{sec:pfrec}.
\end{itemize}

In addition, the user has the freedom to choose the jet clustering algorithm along with its characterizing parameters, as well as minimum threshold for the jet transverse momentum to be stored in the final collection. 
The \DELPHES framework integrates the \FASTJET package~\cite{bib:fastjet} and therefore allows jet reconstruction with the most popular jet clustering algorithms developed so far while keeping track of the constituents. Since most visible objects are reconstructed either as a jet, or as constituents of jets, \DELPHES includes by default in the standard reconstruction sequence a module that automatically removes jets from the event if they have already been reconstructed as isolated electrons, muons or photons. This operation ensures that there is no double-counting of particles in the final-state. Modularity allows this procedure to be easily deactivated if needed by the user.

\subsubsection{$\rm{b}$ and $\rm{\tau}$ jets}\label{sec:btau}

The identification of jets that result from $\rm{\tau}$ decays or the hadronization of heavy flavour quarks --- typically $\rm{b}$ or $\rm{c}$ quarks --- is important in high energy collider experiments.
In \DELPHES a purely parametric approach based on Monte-Carlo generator information has been adopted. 

The algorithm for $\rm{b}$ and $\rm{\tau}$ jet identification proceeds as follows: the jet becomes a potential $\rm{b}$ jet or a $\rm{\tau}$ jet candidate if, respectively, a generated $\rm{b}$ or $\rm{\tau}$ is found within some distance $\Delta R=\sqrt{(\eta^{jet}-\eta^{b,\tau})^2+(\phi^{jet}-\phi^{b,\tau})^2}$ of the jet axis. The probability to be identified as $\rm{b}$ or $\rm{\tau}$ depends on user-defined parameterizations of the $\rm{b}$ and $\rm{\tau}$ tagging efficiency. The user can also specify a mis-tagging efficiency parameterization, that is, the probability that a particle other than $\rm{b}$ or $\rm{\tau}$ be wrongly identified as a $\rm{b}$ or a $\rm{\tau}$. Modularity allows the user to use several $\rm{b}$ and $\rm{\tau}$ tagging algorithms for the same jet collection and to easily implement other tagging algorithms, eventually involving an analysis of the jet constituents. 

\subsection{Missing (transverse) energy and scalar (transverse) energy}\label{sec:metht}

In hadron collider experiments, partons in the initial state having a negligible transverse momentum, the total transverse energy of undetected particles --- the \emph{missing transverse energy} ($E_T^{miss}$) --- can be assessed from the transverse component of the total energy deposited in the detector.
This accounts for example for neutrinos in the standard model but is degraded by the detector resolution, the presence of low momentum looping particles propagating in the forward region and limited acceptance in the forward region.
Another useful quantity is the so-called \emph{scalar transverse energy sum} ($H_T$).
The definition of these two quantities is as follows:
\begin{equation}
\overrightarrow{E_T}^{miss} = - \sum_i \overrightarrow{p_T}(i)\,,
\qquad
H_T = \sum_i |\overrightarrow{p_T}(i)|\,,
\label{eq:metres}
\end{equation}
where the index $i$ runs over the selected input collection. As for the jets, the $E_T^{miss}$ and $H_T$ variables can be computed starting from different input collections. The \emph{Calorimeter }$E_T^{miss}$ and \emph{Calorimeter }$H_T$ variables are estimated by considering only calorimeter towers, while the Particle-Flow $E_T^{miss}$ and Particle-Flow $H_T$ use particle-flow tracks and particle-flow towers as input. 
These quantities can also be calculated using only generator level information. 
Likewise, for $\rm{e^+ e^-}$ collider experiments, \DELPHES is able to compute the total missing energy and the total scalar energy from pure calorimetric information, particle-flow objects or generator level information. 

\section{High-level corrections}\label{sec:hlobj}

So far, we have discussed the procedure in \DELPHES for reconstructing and identifying the most common objects in collider experiments. At this stage, the resulting collections are not yet ready for final analysis. Residual effects such as pile-up contamination and non-uniformity in the energy response need to be corrected for. 
In the following we show how such effects are dealt within \DELPHES.

\subsection{Jet Energy Scale correction}\label{sec:jes}

The average momenta of reconstructed objects do not always match that of their generator-level counterpart. This effect, observed also in real experiments, is particularly explicit in complex objects such as jets where the total smearing is non-trivial due to the clustering procedure, and where parts of the generator-levels components, such as neutrinos, muons and looping particles, are lost.

In \DELPHES, non-composite objects display by construction an average response close to unity.
The energy scale correction is therefore applied only on jets. The user can apply a jet energy scale correction as a function of the reconstructed jet pseudo-rapidity and transverse momentum.

\subsection{Pile-up subtraction}\label{sec:pus}

At the LHC, several collisions per bunch-crossing occur in high luminosity conditions, most of them resulting in a small amount of activity in the detector. 
Due to the elongated shape of the proton bunches constituting the beams, such additional \textit{pile-up events}, take place in a similarly elongated region (called beam spot) around the nominal interaction point. 
In \DELPHES, pile-up interactions are extracted from a pre-generated low-$\rm{Q^2}$ QCD sample. These minimum-bias interactions are randomly placed along the beam axis according to some longitudinal spread that can be set by the user. The actual number of pile-up interactions per bunch-crossing is randomly extracted from a Poisson distribution. 

Pile-up directly affects the performance of jets, $E_T^{miss}$ and isolation. Pile-up interactions are usually identified by means of vertex reconstruction. If such interactions occur far enough from the hard interaction, a precise vertexing algorithm is able to detect them. Combining vertexing and tracking information allows the identification of contaminating charged particles from pile-up. On the other hand, since neutral particles do not produce tracks, neutral pile-up contamination can only be estimated \emph{on average}.  

In real experiments, pile-up mitigation on the missing energy requires the use of advanced techniques which are out of scope in \DELPHES. Therefore no pile-up subtraction is applied on the missing energy variable in \DELPHES. On the other hand, pile-up subtraction is performed on jets and the isolation variable. The procedure involves two steps:

\paragraph{Charged pile-up subtraction}

In \DELPHES the hard scattering occurs at the geometrical centre of the detector. We assume that vertices corresponding to pile-up interactions occurring at a coordinate $\rm{z}$, such that $\rm{|z|>\delta Z_{vtx}}$ can be reconstructed. The parameter $\rm{\delta Z_{vtx}}$ is the spatial vertex resolution of the detector. We assume that pile-up interactions occurring at a coordinate z, such that $\rm{|z|<\delta Z_{vtx}}$ cannot be disentangled from those originating from the high-$\rm{Q^2}$ process. Therefore every charged particle originating from such vertices cannot be subtracted from the event, while every charged particle originating from a vertex positioned at $\rm{|z|>\delta Z_{vtx}}$ can be identified as originating from pile-up, provided that the corresponding track has been reconstructed. For simplicity, in \DELPHES we assume that the track reconstruction efficiency does not vary with the vertex position. If the particle-flow algorithm is being used, the particle-flow tracks identified as originating from pile-up are removed from the list of 4-vector entering the jet clustering and the isolation procedures. 

\paragraph{Residual pile-up subtraction}

Other techniques are needed in order to extract and remove residual contributions: these include particles that are too close to the hard interaction vertex to be identified as pile-up products with tracking information, charged particles that failed track reconstruction (or outside the tracker volume) and neutral particles. In \DELPHES we have opted for the Jet Area method~\cite{bib:jetarea1,bib:jetarea2}. This approach, widely used in present collider experiments, allows the extraction of an average contamination density $\rho$ on an event-by-event basis. In practice, this is performed in \DELPHES with the help of the \FASTJET package.

The pile-up density $\rho$, can then be used to correct observables that are sensitive to the residual contamination, the jet energies and the isolation variable (defined in equation~(\ref{eq:isolation})). 
In the presence of residual pile-up contamination, these two quantities are corrected in the following way:
\begin{eqnarray}
p_{jet}&\rightarrow&p_{jet} - \rho \cdot A_{jet},\\
I(P)&\rightarrow&I(P)-\frac{\rho \cdot \pi R^2}{p_T(P)},
\end{eqnarray}
where $A_{jet}$ is the jet area estimated via the \FASTJET package, and $R$ is the diameter of the isolation cone. 

The separate treatment of the charged and the neutral pile-up components is particularly effective if combined with the particle-flow reconstruction approach. As already mentioned, particle-flow tracks that are not associated with the hard interaction as well as their corresponding calorimeter deposit can be removed from the input 4-vectors that enter the jet clustering procedure, provided that the particle-flow algorithm is switched on. The neutral energy offset can then be estimated with the Jet Area method. If no tracking information is available (for \emph{Calorimeter Jets} for instance), one can simply estimate the global event pile-up contribution with the Jet Area method. 

\section{Validation}\label{sec:validation}

The simulation and reconstruction in \DELPHES has to be validated by comparing the resolution of the output objects to the resolutions of real experiments. We chose to validate \DELPHES against the two major multipurpose collider experiments presently in operation, CMS~\cite{bib:cmstdr} and ATLAS~\cite{bib:atlastdr}. Only the performance of high-level objects such as electrons, muons, photons, jets and $E_T^{miss}$, is discussed here. All the Monte-Carlo samples used for the validation are produced with the \MG event generator~\cite{bib:madg} and hadronized with \PYTHIA~\cite{bib:pythia}. In order to properly account for tree-level higher order QCD contributions, the $k_T$-MLM matching procedure was applied~\cite{bib:mlm}. Events are then processed by \DELPHESTZE with specific CMS and ATLAS configurations.~\footnote{A default CMS and ATLAS configuration card is included within each \DELPHES release.} The nominal detector resolutions are used for CMS~\cite{bib:cmstdr} and ATLAS~\cite{bib:atlastdr}. The ECAL granularity is set equal to the HCAL granularity for both detectors.

\subsection{Charged leptons and photons}\label{sec:chlepphoval}

\begin{figure}[tbp]
\centering
\begin{minipage}{.5\textwidth}
  \centering
  \includegraphics[width=1\linewidth]{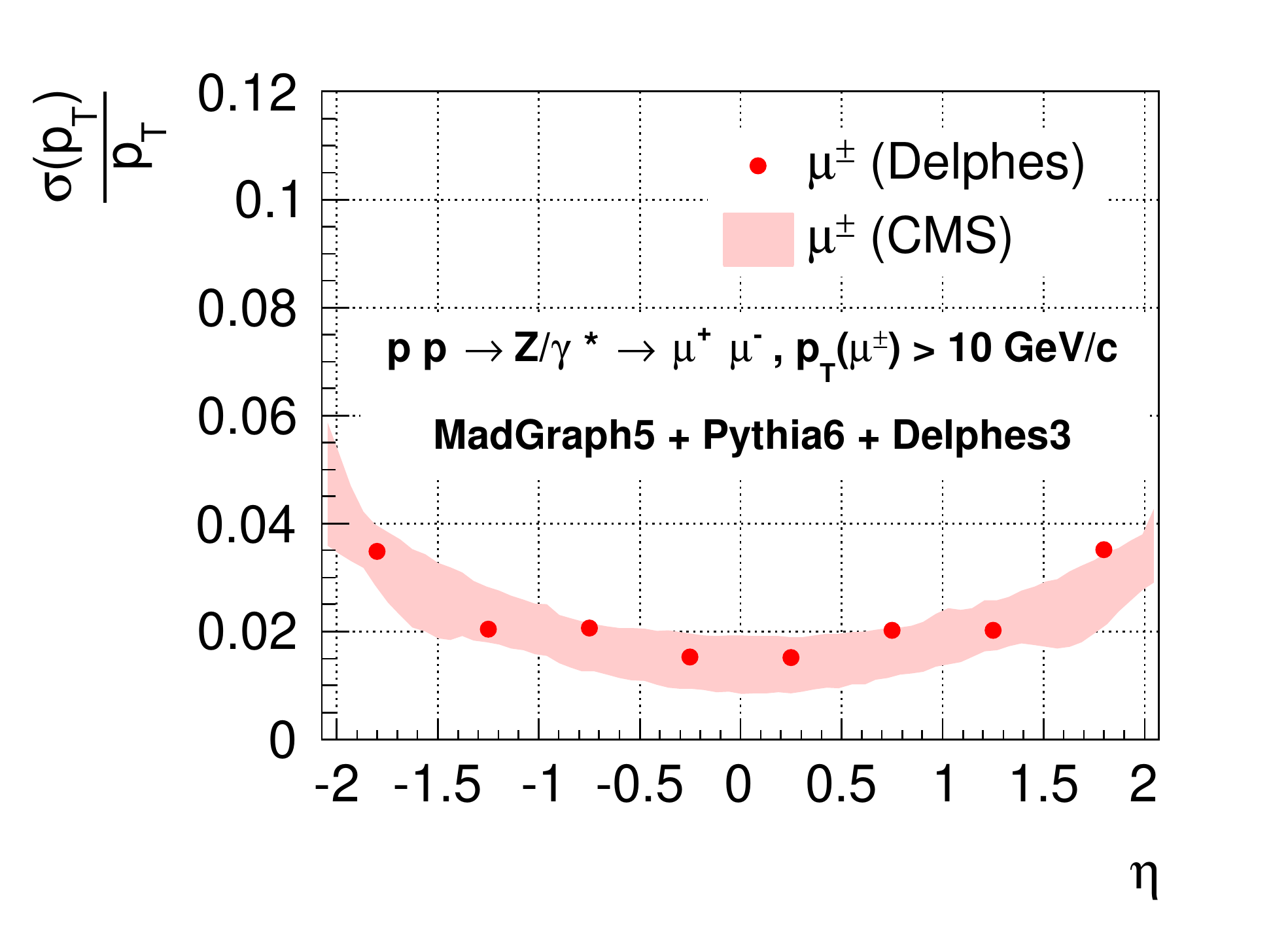}
\end{minipage}%
\begin{minipage}{.5\textwidth}
  \centering
  \includegraphics[width=1\linewidth]{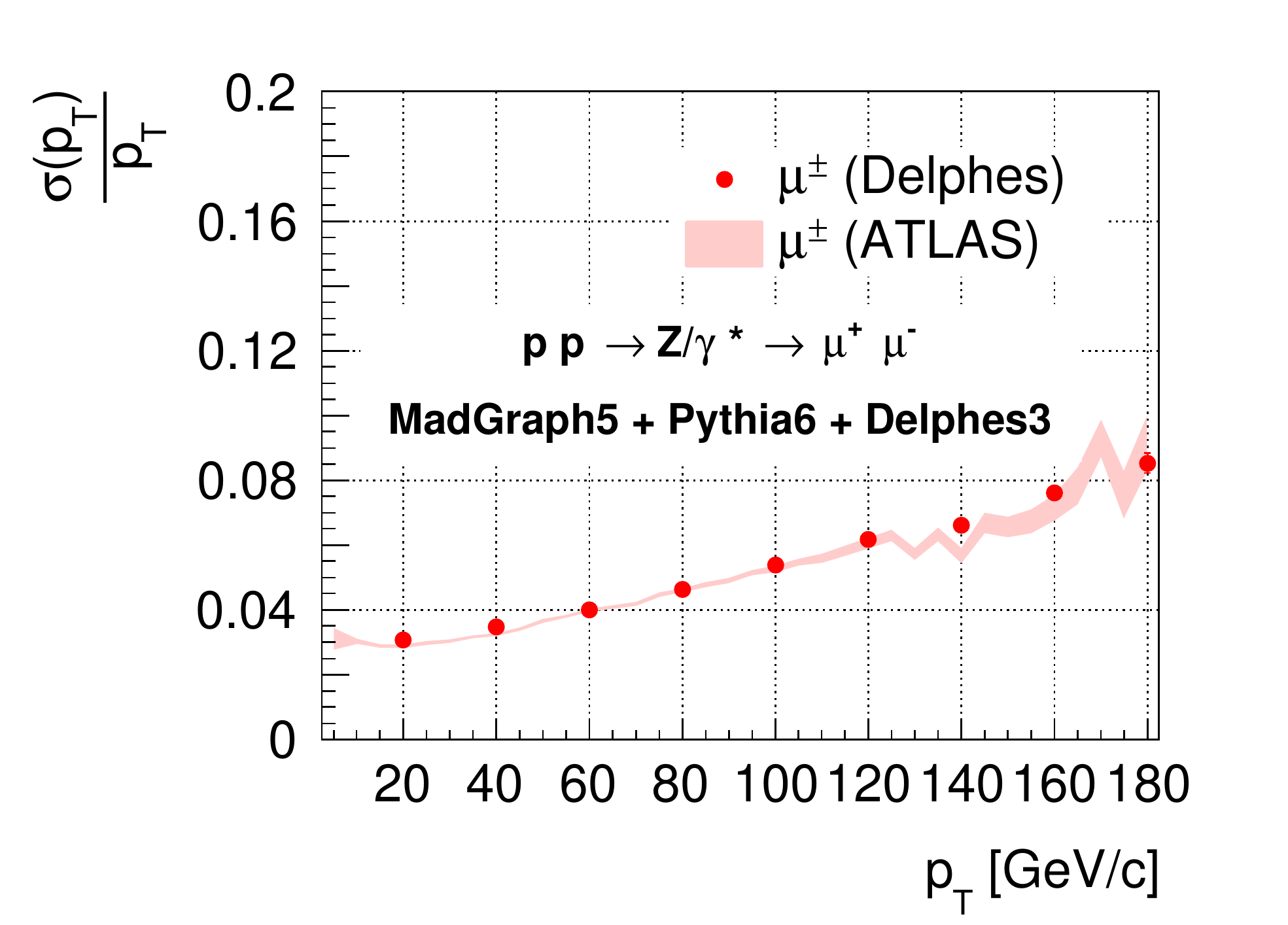}
\end{minipage}
\caption{Left: muon $p_T$ resolution as function of $\eta$ for \DELPHES and CMS. Right: muon $p_T$ resolution as function of $p_T$ for \DELPHES and ATLAS. The resolution obtained with \DELPHES is shown with circular dots and the corresponding statistical uncertainty is shown with vertical error bars, mostly hidden by the dots. For the CMS comparison (left) the band represents the overall (systematic+statistical) uncertainty resulting from the measurement of the muon momentum resolution in CMS data~\cite{bib:muoncms}. For ATLAS (right) the band represents the statistical uncertainty on the resolution obtained in simulation~\cite{bib:muonatlas}.}  
 \label{fig:muocms}
\end{figure}

\begin{figure}[tbp]
\centering
  \includegraphics[width=0.8\linewidth]{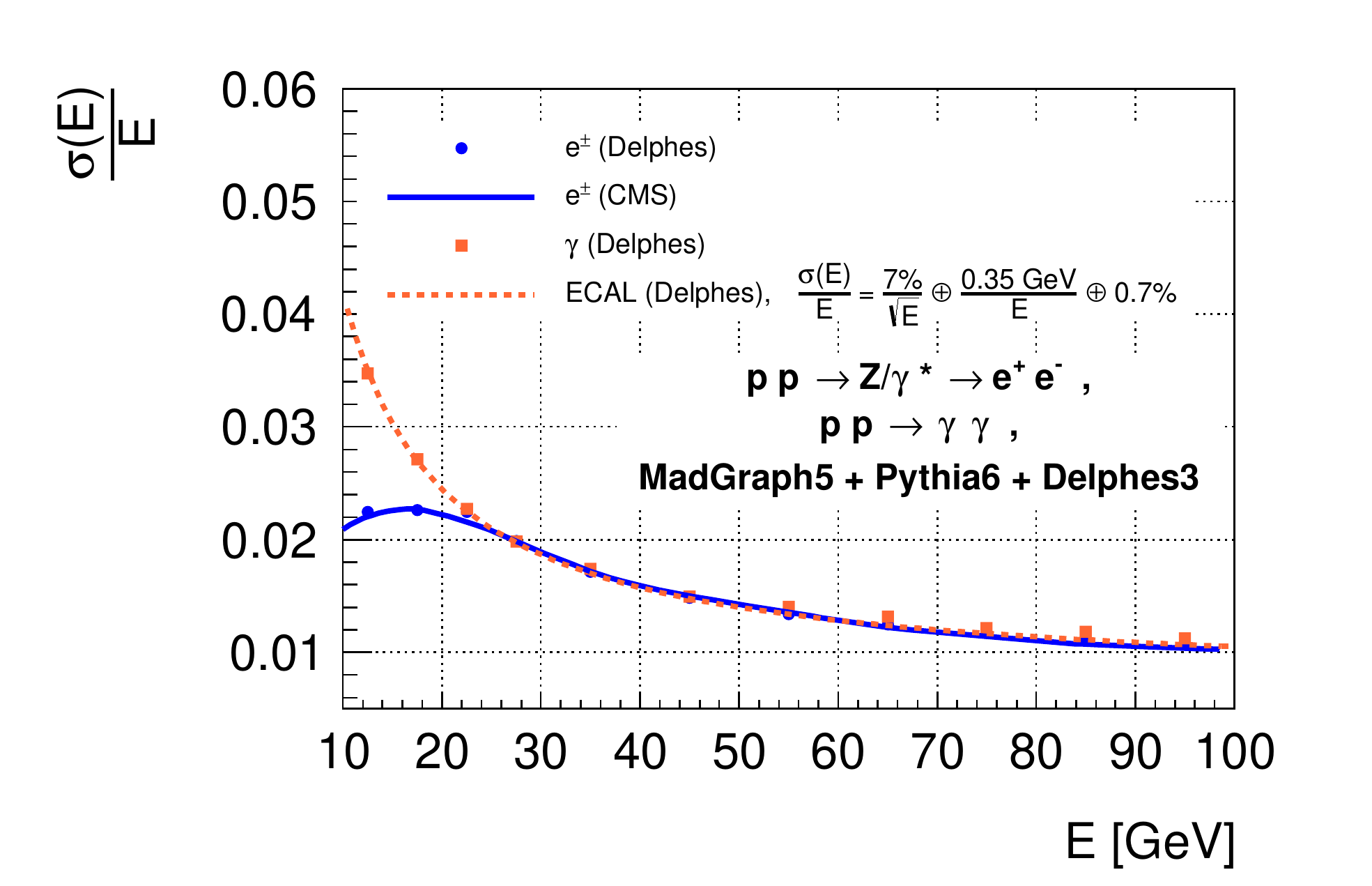}
\caption{Electron and photon energy resolution as a function of the energy for a CMS-like detector. The CMS gaussian electron resolution is from~\cite{bib:elecms}. At high energy, the electron and photon resolutions are driven by ECAL and are therefore identical. At low energy, the electron resolution is largely driven by the superior tracking resolution.}
\label{fig:egamma}
\end{figure}

Electrons and muons are generated from two independent $\rm{p p \rightarrow Z/\gamma* \rightarrow e^+ e^-}$ and $\rm{p p}$ $\rm{\rightarrow}$ $\rm{Z}$/$\rm{\gamma*}$ $\rightarrow$ $\rm{\mu^+ \mu^-}$ samples, while photons are obtained from a $\rm{p p \rightarrow \gamma \gamma}$ sample. The resolution is computed as follows. For each generated $\rm{e^{\pm}}$ ($\rm{\mu^{\pm}}$, $\rm{\gamma}$), we look for the reconstructed $\rm{e^{\pm}}$ ($\rm{\mu^{\pm}}$, $\rm{\gamma}$) candidate with the smallest $\Delta R=\sqrt{(\eta^{rec}-\eta^{gen})^2+(\phi^{rec}-\phi^{gen})^2}$. If $\Delta R$<0.2, the generated particle is paired with a reconstructed isolated particle. The energy resolution is computed, for each bin, as the Gaussian variance of the distribution of the ratio $(E^{gen}-E^{rec})/E^{gen}$ (see for instance figure~\ref{fig:egamma}). Alternatively, the transverse momentum resolution is computed as the variance of the ratio $(p_T^{gen}-p_T^{rec})/p_T^{gen}$, as shown in figure~\ref{fig:muocms}. 

A comparison of the muon $p_T$ resolution obtained with \DELPHES and the CMS~\cite{bib:muoncms} and ATLAS~\cite{bib:muonatlas} detectors is shown in figure~\ref{fig:muocms}. The agreement is good for both.

In figure~\ref{fig:egamma} the electron and photon energy resolution are shown. For comparison the electron gaussian energy resolution from CMS~\cite{bib:elecms} is also shown. The electron resolution agrees well between CMS and \DELPHES. As an illustration, we show also the nominal ECAL resolution in \DELPHES. At high energies the electron and photon resolutions match perfectly the ECAL resolution. At low energies, the electron resolution is driven by the tracking resolution.

\subsection{Jets}\label{sec:jetval}

The validation of jets is performed on QCD events. The jet energy resolution is obtained in a similar way as explained in section~\ref{sec:chlepphoval} by matching reconstructed and generated jets. For both CMS and ATLAS jets are clustered with the anti-$k_T$~\cite{bib:antikt} algorithm with a cone parameter $\Delta R=0.5$ and $\Delta R=0.6$ respectively. 

\begin{figure}[tbp]
\centering
\begin{minipage}{.5\textwidth}
  \centering
  \includegraphics[width=1\linewidth]{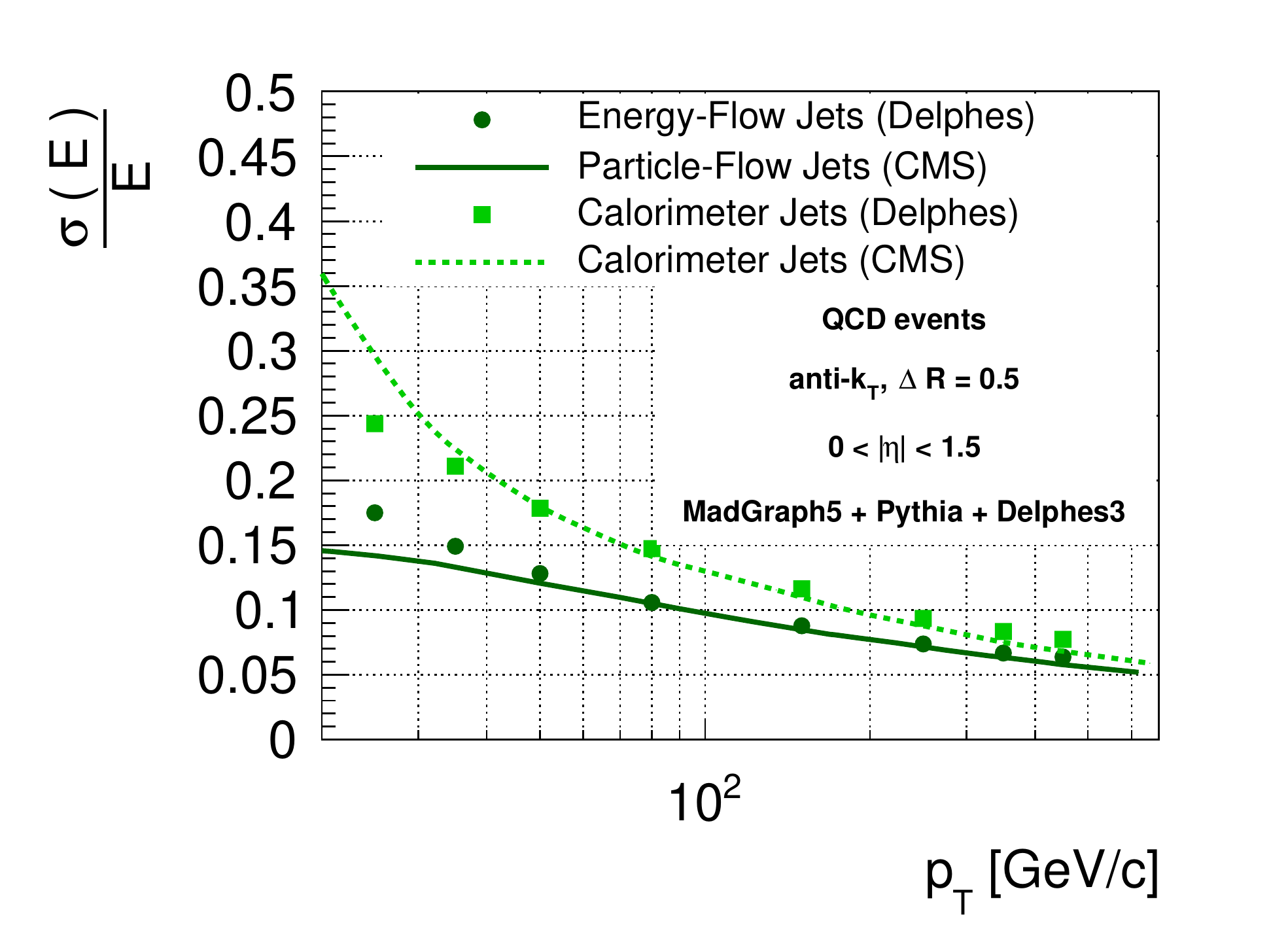}
\end{minipage}%
\begin{minipage}{.5\textwidth}
  \centering
  \includegraphics[width=1\linewidth]{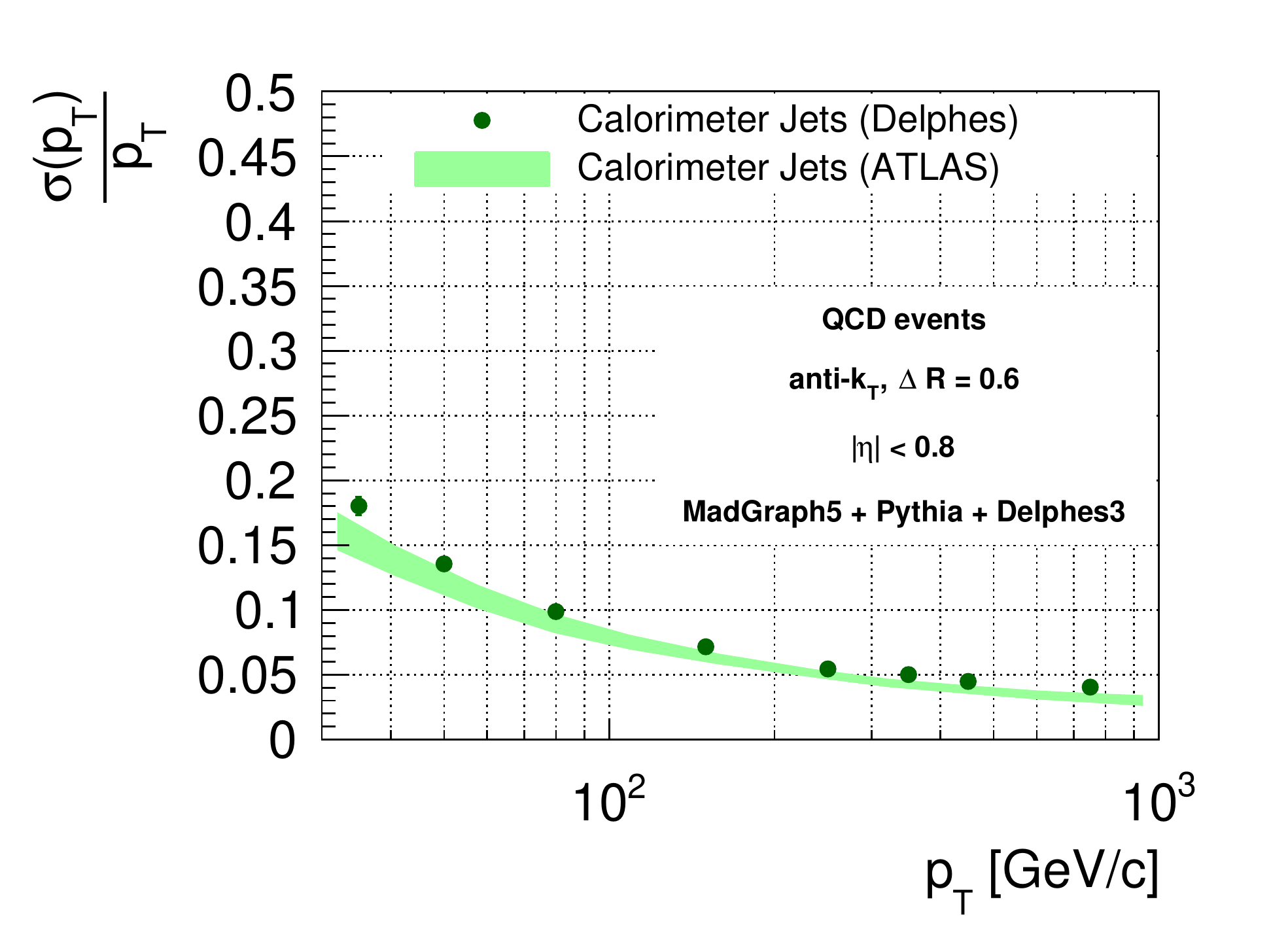}
\end{minipage}
\caption{Left: Comparison of the jet energy resolution of Particle-Flow and Calorimeter jets in CMS~\cite{bib:pflow} and \DELPHES.
Right: Comparison of the jet energy resolution of Calorimeter jets in \DELPHES and ATLAS~\cite{bib:jetatlas}. The band represents the resolutions obtained with different methods in the ATLAS simulation.}
  \label{fig:jetcms}
\end{figure}

In figure~\ref{fig:jetcms} (left) a comparison between CMS and \DELPHES resolutions is shown for Calorimeter Jets and Particle-Flow Jets. The ECAL and HCAL calorimeter resolutions have been set to the actual CMS resolutions. Both approaches show a good agreement with CMS results~\cite{bib:pflow}. In particular, the agreement is perfect at medium and high $p_T$ values ($p_T>40$ Gev/c). There is a significant discrepancy for $20<p_T<30$ GeV/c that is not understood. However this can hardly affect physics analyses, where mostly jets with $p_T$ > 30 GeV/c are considered. For the ATLAS comparison only Calorimeter Jets resolutions are shown (figure~\ref{fig:jetcms} (right)). Also in this case, \DELPHES reproduces with good accuracy the ATLAS results~\cite{bib:jetatlas}.

\subsection{Missing Transverse Energy}\label{sec:metval}

The $E_T^{miss}$ performance is validated both on events with neutrinos (real $E_T^{miss}$) and without neutrinos (fake $E_T^{miss}$) in the final state. The fake $E_T^{miss}$ validation is performed in the presence of pile-up.

\begin{figure}[tbp]
\centering
\begin{minipage}{.5\textwidth}
  \centering
  \includegraphics[width=1\linewidth]{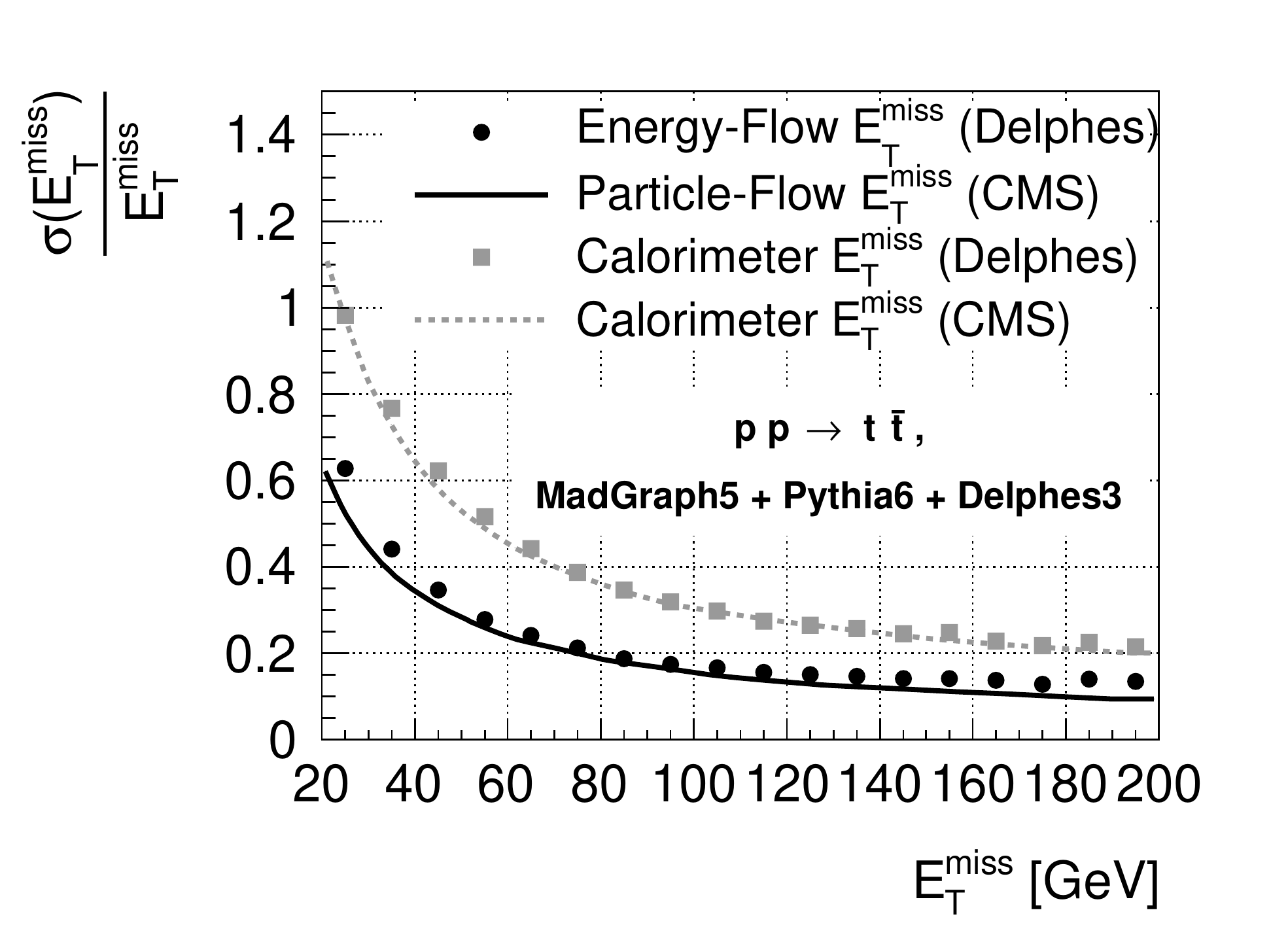}
\end{minipage}%
\begin{minipage}{.5\textwidth}
  \centering
  \includegraphics[width=1\linewidth]{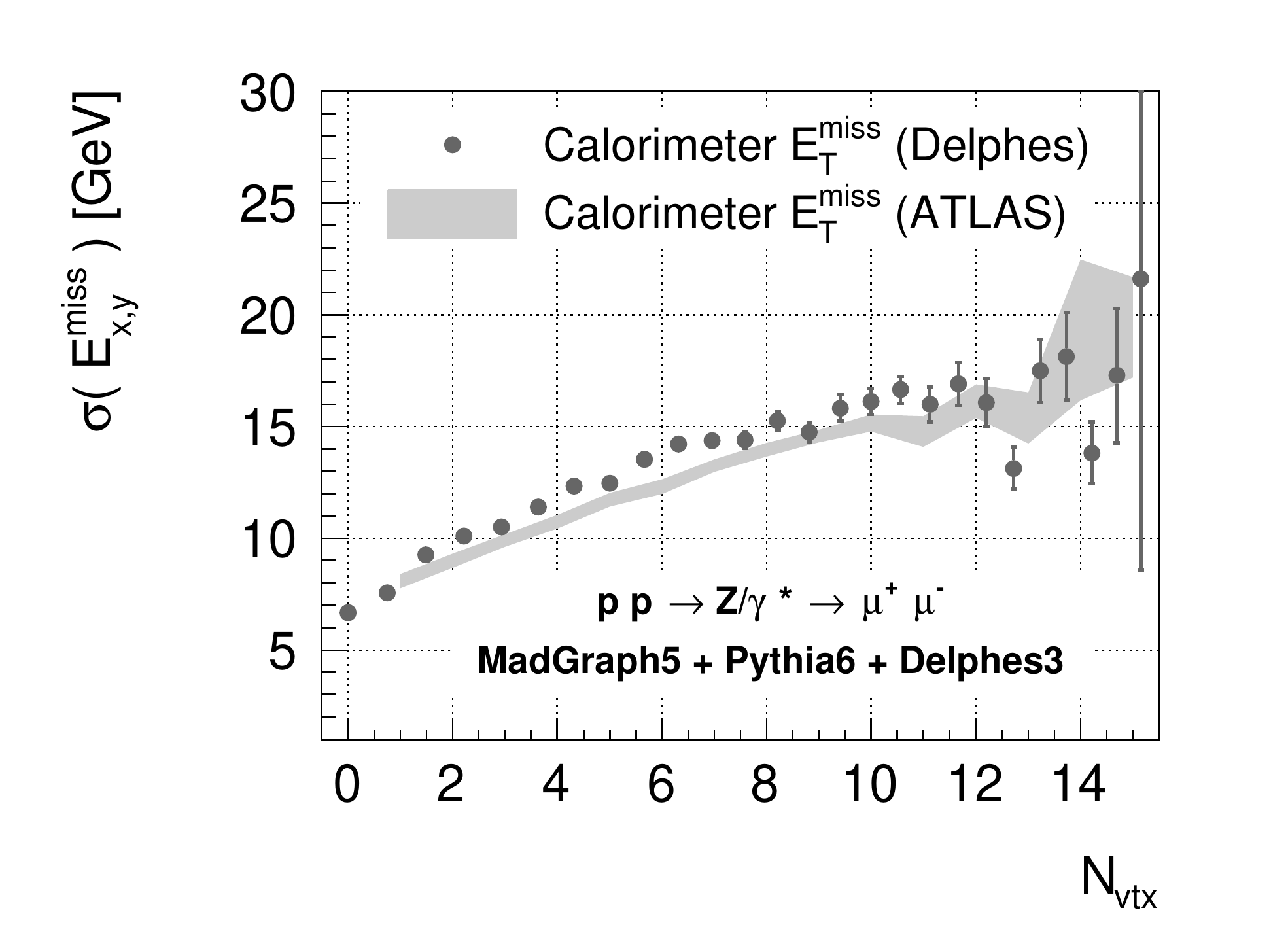}
\end{minipage}
\caption{Left: Particle-Flow $E_T^{miss}$ and Calorimeter $E_T^{miss}$ resolution in \DELPHES and CMS~\cite{bib:pflow}. 
Right: $E_{x,y}^{miss}$ resolution in \DELPHES and ATLAS as a function of the number of reconstructed primary vertices~\cite{bib:metatlas}. The grey band represents the discrepancy between the ATLAS simulation and data.}
  \label{fig:metres}
\end{figure}

Inclusive top pair events are used for testing the real $E_T^{miss}$ performance. The resolution is computed as usual by comparing the $E_T^{miss}$ obtained with calorimeter towers, or particle-flow candidates with the sum of neutrino transverse momenta at parton level. The resolution, as a function of the true $E_T^{miss}$ in the event, is shown in figure~\ref{fig:metres} (left). Both for the Calorimeter and Particle-Flow $E_T^{miss}$ the agreement between \DELPHES and CMS~\cite{bib:pflow} is good.

The fake $E_T^{miss}$ performance is asserted by means of a $\rm{Z/\gamma}$* $\rm{\rightarrow}$ $\rm{\mu^+\mu^-}$ sample. Following the approach of the ATLAS collaboration~\cite{bib:metatlas}, we select events by requiring the di-muon invariant mass to be compatible with the Z boson mass and we reject events where at least one jet with $p_T~>~20$ GeV has been reconstructed. The resolution of the x and y components of the $E_{T}^{miss}$ as a function of the number of reconstructed primary vertices is shown in figure~\ref{fig:metres} (right).

Since no vertex reconstruction is performed in \DELPHES, the number of reconstructed vertices is simply obtained by rescaling the number of generated pile-up interactions by a pile-up dependent factor. This factor accounts for the vertex reconstruction efficiency in the presence of pile-up. The vertex reconstruction efficiency is assumed to decrease linearly as a function of the number of true pile-up interactions. It varies from 75\% when only the hard-scattering occured (pile-up $\approx$ 0) to 50\% at high pile-up conditions ($\approx$ 40). These numbers have been extracted from~\cite{bib:vtxatlas}.

\section{Use cases}\label{sec:usecase}

In order to illustrate the \DELPHES fast-simulation with concrete examples, two use cases are developed in the following.
In the first example, the mass of the top quark is reconstructed in semi-leptonic $\rm{t\bar{t}}$ events. 
The performance of the reconstruction and selection is compared with the literature.
In the second example, the impact of the presence of pile-up on a typical vector boson fusion Higgs analysis workflow is illustrated.
Both examples are distributed as part of the \DELPHES releases and are meant to be easy to understand. The following results have been obtained with \DELPHESTZE.

\subsection{Top Quark mass}\label{sec:top}

In modern collider experiments at high energy, $\rm{t\bar{t}}$ events are among the most copious signatures observed in the detectors.
When one top quark decays leptonically and the other hadronically, the signature is characterized by one lepton, missing transverse energy and four jets, two of them originating from the fragmentation of b quarks. Moreover, at the LHC, about 50\% of events have extra hard jets coming from initial or final state radiation.
Following the semi-leptonic $\rm{t\bar{t}}$ analysis described in ref.~\cite{bib:ttbarcms} we focus on the mass of the hadronically-decaying top quark.

The $\rm{t\bar{t}}$+jets sample has been generated with \MG at a centre of mass energy $\sqrt{s}=7$ TeV and \PYTHIA was used for parton shower and hadronization. Backgrounds are not considered here. 
The reconstruction has been performed via \DELPHES using the detector configuration designed to mimic the performance of the CMS detector. 

Following the CMS approach, we select events with exactly one isolated lepton (electron or muon) with $p_T>30$ GeV/c and $|\eta|<2.1$. In addition, we require at least four particle-flow jets with $p_T>30$ GeV/c and $|\eta|<2.4$. The anti-$k_T$~\cite{bib:antikt} algorithm with a parameter $R=0.5$ was used for jet clustering. 
Among the selected jets, at least two must be tagged as originating from the hadronization of a $\rm{b}$ quark ($\rm{b}$-tagged) and at least two must be identified as \emph{light jets} (i.e. fail the b-tagging criterion). 
The b-tagging efficiency parameterization has been extracted from~\cite{bib:btagcms}. The signal efficiency for this selection is 2.8\%, compared to 2.3\% in the CMS analysis, showing a reasonable agreement between \DELPHES and CMS. Given the high jet multiplicity, the signal selection is extremely sensitive to changes in requirements that can affect the jet selection. 
 
Since selected events contain two b-jets ($b_1$ and $b_2$) and two light jets ($j_1$ and $j_2$), among the four leading jets two choices are possible for reconstructing the hadronic top mass: ($b_1, j_1, j_2$) and ($b_2, j_1, j_2$). Following the CMS definition, each of the two possible assignments can be classified as:
\begin{itemize}
\item \emph{unmatched}, if there is at least one of the four observed leading jets that does not match any parton from the decay of either of top quarks.
\item \emph{wrong permutation}, if the four leading jets match with the four partons but the assignment of the reconstructed b-jet with the b parton originating from the hadronically decayed top leg is wrong,
\item \emph{correct permutation} if all the jet-parton assignments are correct.
\end{itemize}
 
\begin{figure}[tbp]
 \begin{center}
  \includegraphics[width=\textwidth]{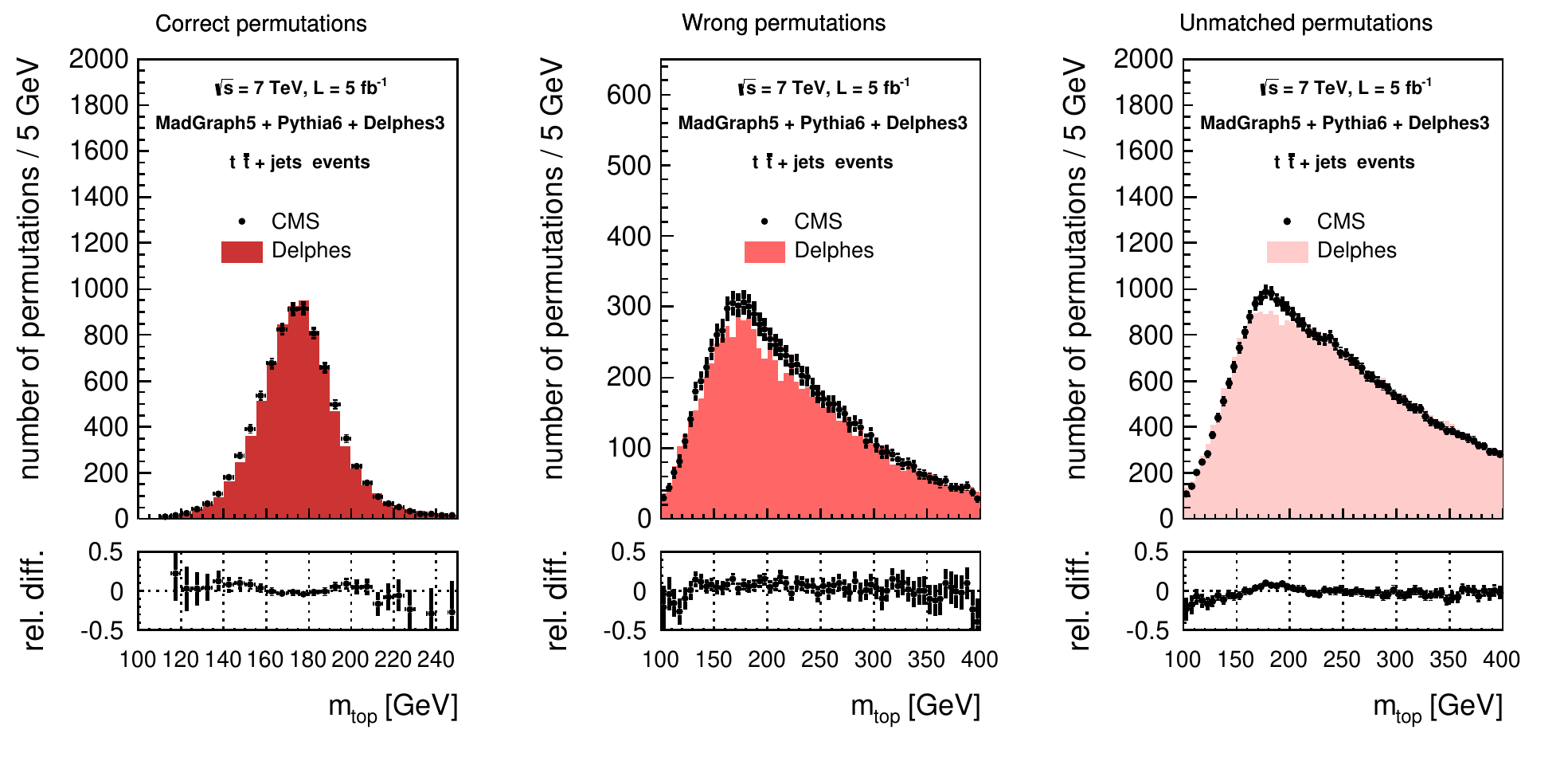}
  \caption{Reconstructed hadronic top mass distributions for the \emph{correct assignments} (left), \emph{wrong assignments} (centre) and \emph{unmatched permutations} (right). The \DELPHES distributions are normalized to the CMS yield. The CMS contributions are taken from ref.~\cite{bib:ttbarcms}.}
  \label{fig:ttbar_perm}
 \end{center}
\end{figure}

\begin{figure}[tbp]
\centering
\begin{minipage}{.5\textwidth}
  \centering
  \includegraphics[width=1\linewidth]{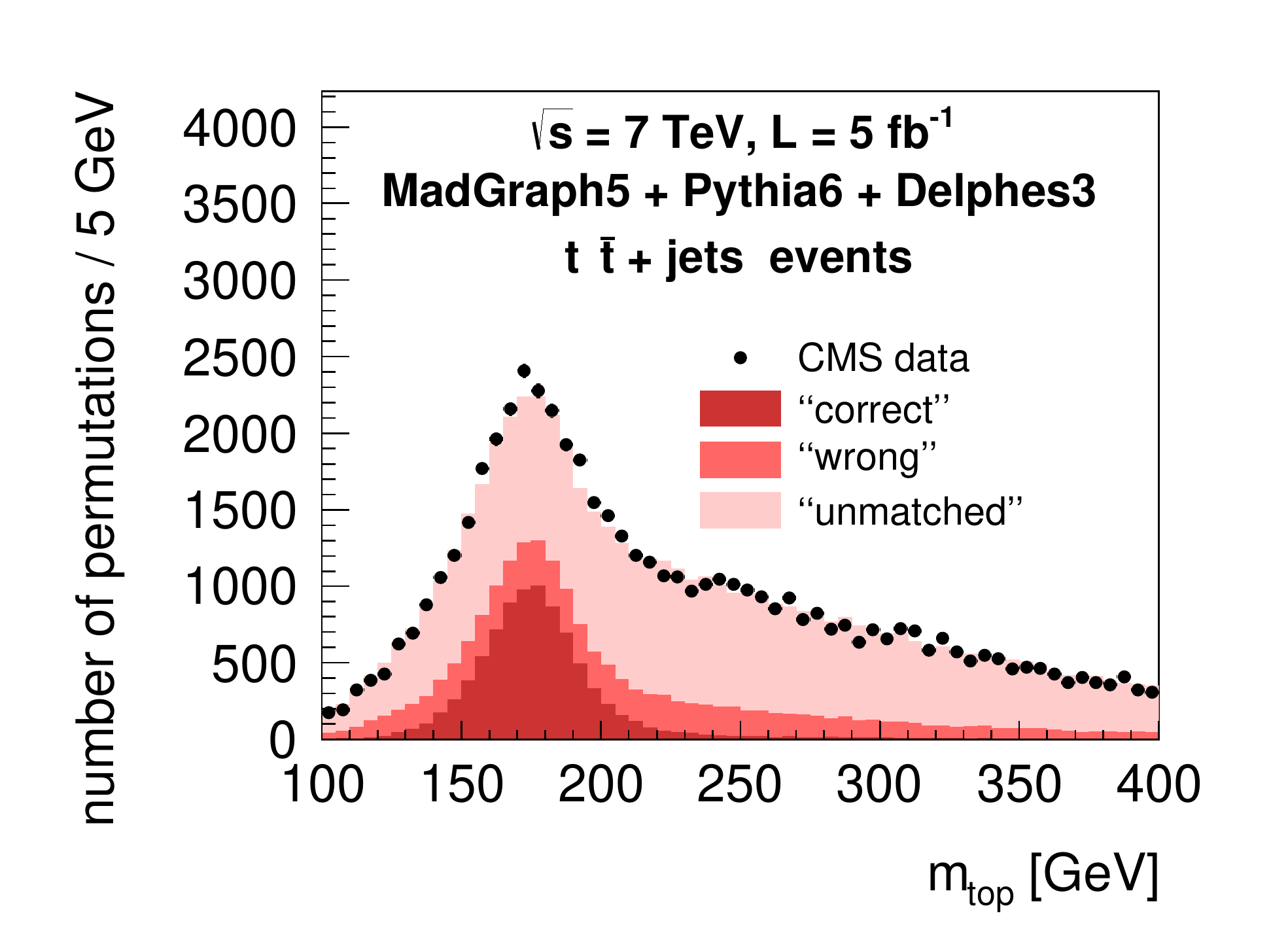}
\end{minipage}%
\begin{minipage}{.5\textwidth}
  \centering
  \includegraphics[width=1\linewidth]{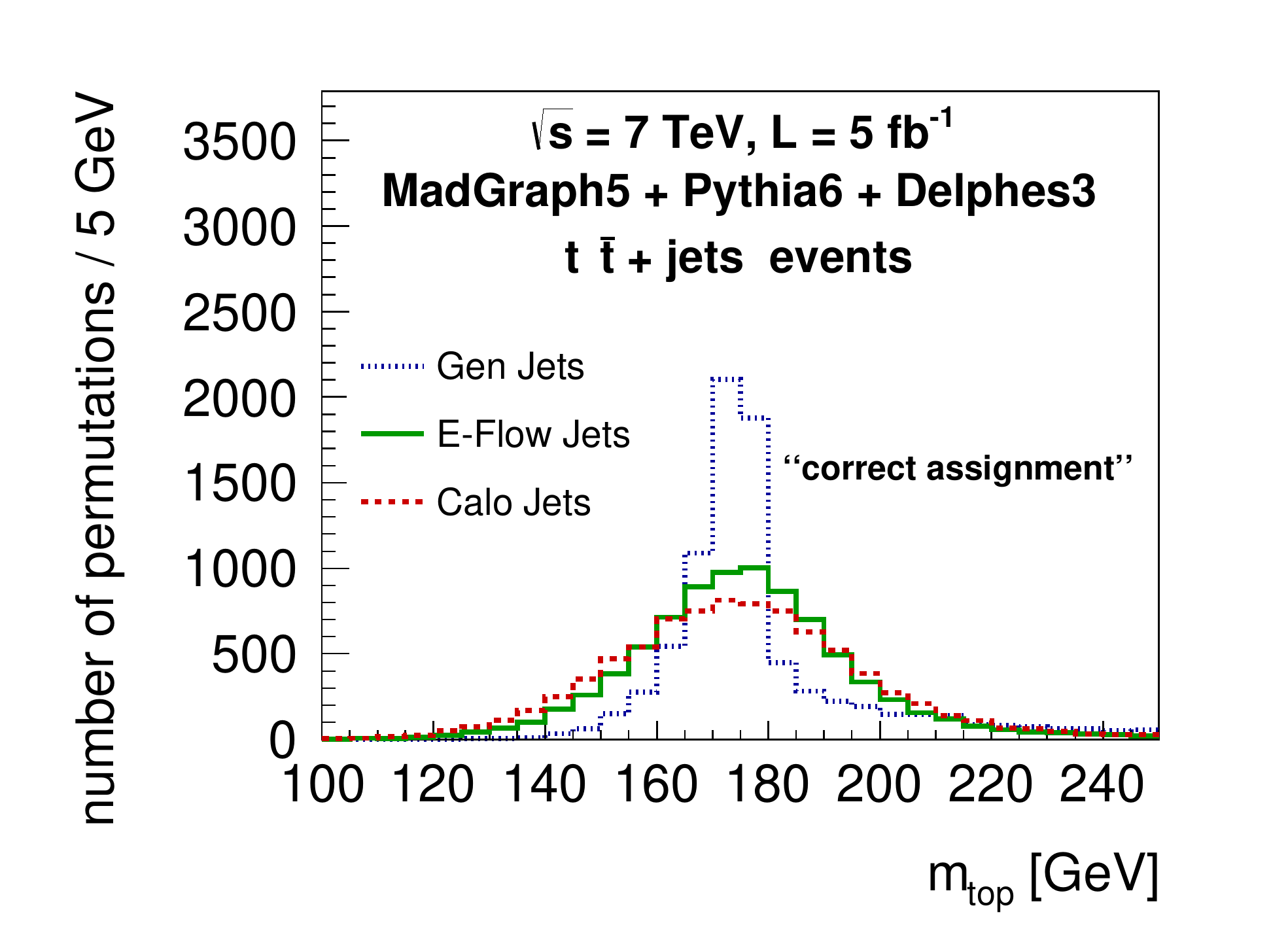}
\end{minipage}
\caption{Left: Reconstructed hadronic top mass for all possible assignment categories. The \DELPHES distribution is normalized to the CMS yield. The CMS contributions are taken from Ref.~\cite{bib:ttbarcms}. Right: Reconstructed hadronic top mass distribution for the \emph{correct assignments} only. The distribution is shown for \emph{Generated Jets}, \emph{Particle-Flow Jets} and \emph{Calorimeter Jets}.}
  \label{fig:mtop_res}
\end{figure}

The relative fraction of each permutation category has been compared with the fractions obtained by the CMS collaboration, showing a good agreement (see table~\ref{tab:eff_cms}). 
The top quark mass distributions obtained with \DELPHES and CMS for the three permutation categories are shown in figure~\ref{fig:ttbar_perm} and~\ref{fig:mtop_res}, left.
The \DELPHES distributions are normalized to the CMS total number of events.
Overall the shapes and relative contributions corresponding to the three categories are well reproduced by \DELPHES. 

\begin{table}[btp]
 \begin{center}
  \begin{tabular}{|l|c|c|}
  \hline
            & CMS     & \DELPHES \\
  \hline
  correct   & 15.5 \% & 15.8 \% \\
  \hline
  wrong     & 17.4 \% & 16.5 \% \\
  \hline
  unmatched & 67.1 \% & 67.7 \% \\
  \hline
  \end{tabular}
  \caption{Fractions of permutations for each type of assignment in the top-mass range $100$ to $400$ GeV for CMS and \DELPHES.}
  \label{tab:eff_cms}
 \end{center}
\end{table}

For the sake of illustration, the reconstructed hadronic top mass using correct permutations only is shown in figure~\ref{fig:mtop_res} (right) using three different jet collections: \emph{Generated Jets}, \emph{Calorimeter Jets} and \emph{Particle-Flow Jets}, defined in section~\ref{sec:jetval}. We observe, as expected, a narrow peak when using \emph{Generated Jets} and wider peaks when using \emph{Particle-Flow Jets} or \emph{Calorimeter Jets}. 
This illustrates the need for using realistically reconstructed objects rather that hadron-level quantities in prospective phenomenological studies.

\subsection{Higgs Production via Vector Boson Fusion with pile-up}\label{sec:vbfh}

The observation of a Higgs particle decaying to a $\rm{\bbbar}$ pair, produced via Vector Boson Fusion, can be useful in order to constrain the VVH and bbH couplings in the standard model. 
The signal being characterized by a fully hadronic final state, the favorable branching ratio of the $\rm{H\rightarrow\bbbar}$ decay is heavily counterbalanced by the presence of large QCD backgrounds at the LHC. 
Moreover, the presence of pile-up is expected to have a large impact on the jet reconstruction and on the \emph{rapidity gap} requirement. 
These aspects make this search very challenging, especially at high luminosity, and an ideal playground for testing \DELPHES capabilities.

\begin{figure}[tbp]
 \begin{center}
  \includegraphics[width=0.7\textwidth]{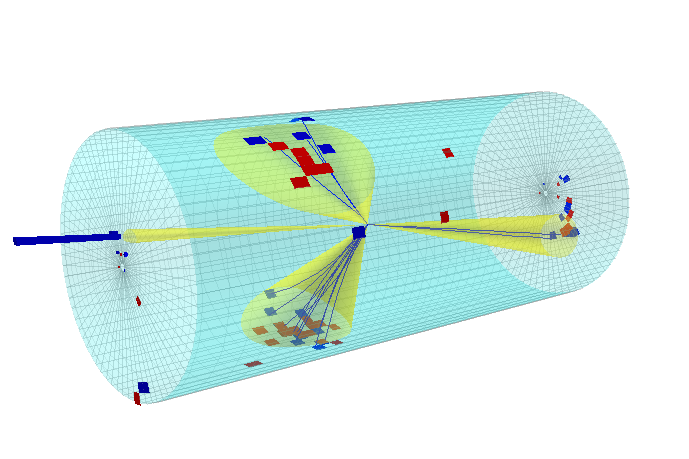}
  \caption{A typical Vector Boson Fusion $\rm{H \rightarrow \bbbar}$ event shown with the \DELPHES event display. The event principally contains two forward jets with a large rapidity-gap and two central (b)-jets.}
  \label{fig:evdisplay}
 \end{center}
\end{figure}

The signal signature is characterized by the presence of two highly energetic jets at high rapidity. 
Since no color flow is exchanged between the two jets, little hadronic activity is expected in the central part of the detector, besides the Higgs decay products. A typical signal event is shown in figure~\ref{fig:evdisplay}, with the help of the \DELPHES event display. In large pile-up scenarios, additional jets might be reconstructed in the central region of the detector, hence spoiling the sensitivity of this search. 

Both the signal and background samples have been generated with \MG~\cite{bib:madg} at a centre of mass energy $\sqrt{s}=14$ TeV. Only the main irreducible $\rm{\bbbar+jets}$ background was considered. Events have been showered and hadronized via \PYTHIA~\cite{bib:pythia}. Detector simulation and event reconstruction has been performed with \DELPHES. Pile-up events originate from a Minimum Bias sample generated with \PYTHIAEIGHT~\cite{bib:pythia8}.

Jets are the only relevant objects to be considered for this analysis. In order to fully explore the pile-up mitigation potential in \DELPHES, particle-flow jets are used for this analysis. The anti-$k_T$~\cite{bib:antikt} algorithm with a parameter $R=0.5$ was adopted for the jet clustering of particle-flow input objects. In the central region of the detector, where tracking information is available, charged particles originating from pile-up are removed from the particle-flow object collection before the jet clustering procedure. The residual pile-up contamination, originating mainly from neutrals, is estimated via the Jet Area method (see section~\ref{sec:pus}). The pile-up density $\rho_{cen}$, used for the residual subtraction, has been estimated in the central part of the detector only. In the forward region, where no tracking information is available, the total (charged+neutral) pile-up contamination density $\rho_{fwd}$ is computed and then used to correct the jet energy.

The following event selection was applied:

\begin{enumerate}
\item at least 4 jets with $p_T$ > 80, 60, 40, 40 GeV respectively, of which at least 2 b-tagged jets ($b_1$, $b_2$) and at least 2 light jets ($j_1$, $j_2$) , 
\item $\Delta \eta_{j_1j_2}$ > 3, $\eta_{j_1}\times\eta_{j_2}$ < 1, $m_{j_1j_2}$ > 500 GeV, no light jets between $j_1$ and $j_2$,
\item 100 < $m_{b_1b_2}$ < 200 GeV. 
\end{enumerate}

The three selection steps are aimed at increasing the signal-to-background ratio. Selection criterion (1) addresses the threshold of the jet momenta. Jets are typically expected to be softer in QCD backgrounds than in the signal, especially the b-jets that, in the signal case, originate from a heavy resonance. Selection (2) addresses specifically the difference in topology between signal and background. The two hardest light jets are required to have a large rapidity gap, a high dijet invariant mass, and no hadronic activity in between, besides the two b's originating from the Higgs decay. Selection (3) further increases the signal purity by requiring a $\rm{\bbbar}$ invariant compatible with the Higgs resonance. 

\begin{figure}[tbp]
\centering
\begin{minipage}{.5\textwidth}
  \centering
  \includegraphics[width=1\linewidth]{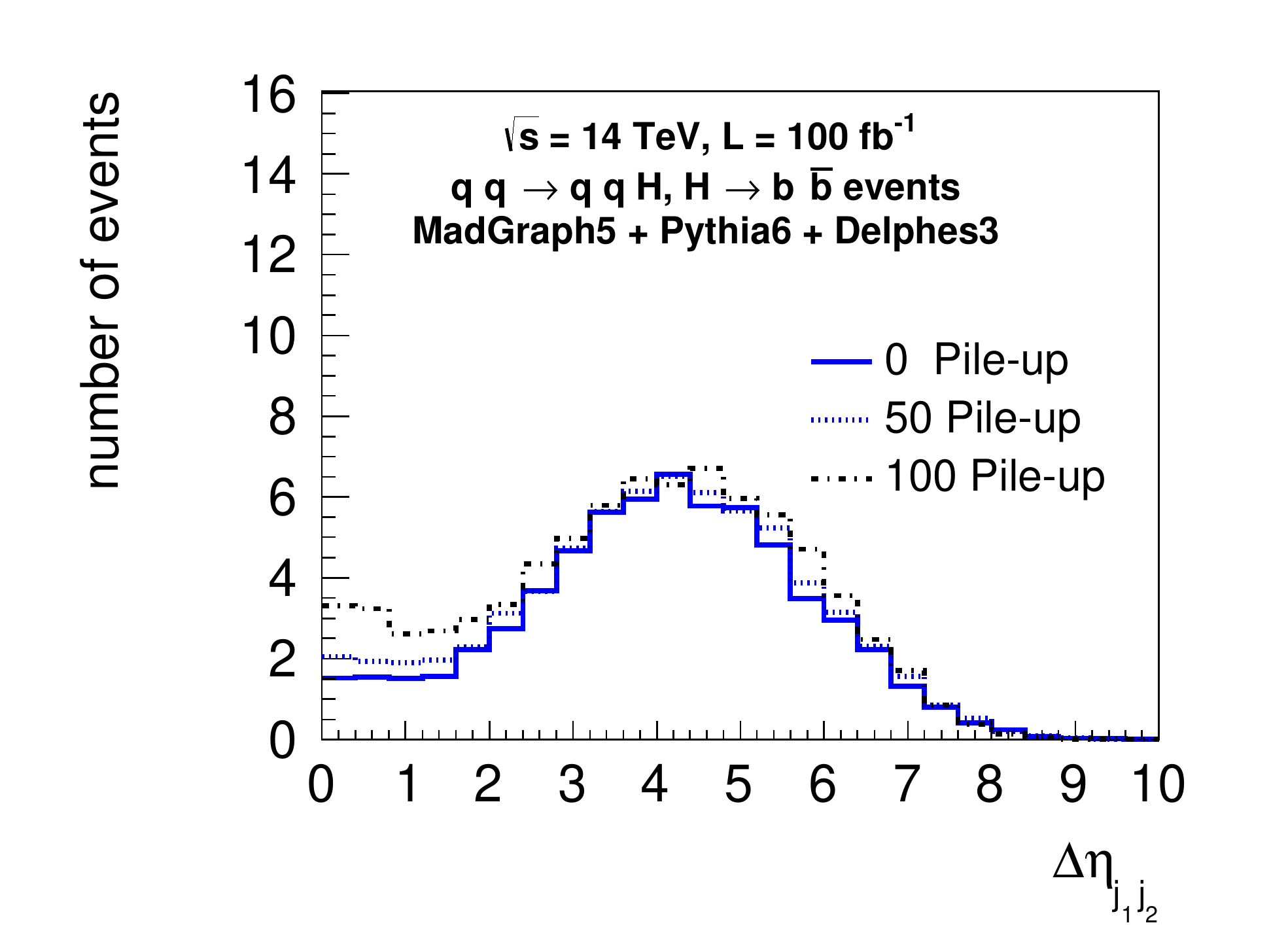}
\end{minipage}%
\begin{minipage}{.5\textwidth}
  \centering
  \includegraphics[width=1\linewidth]{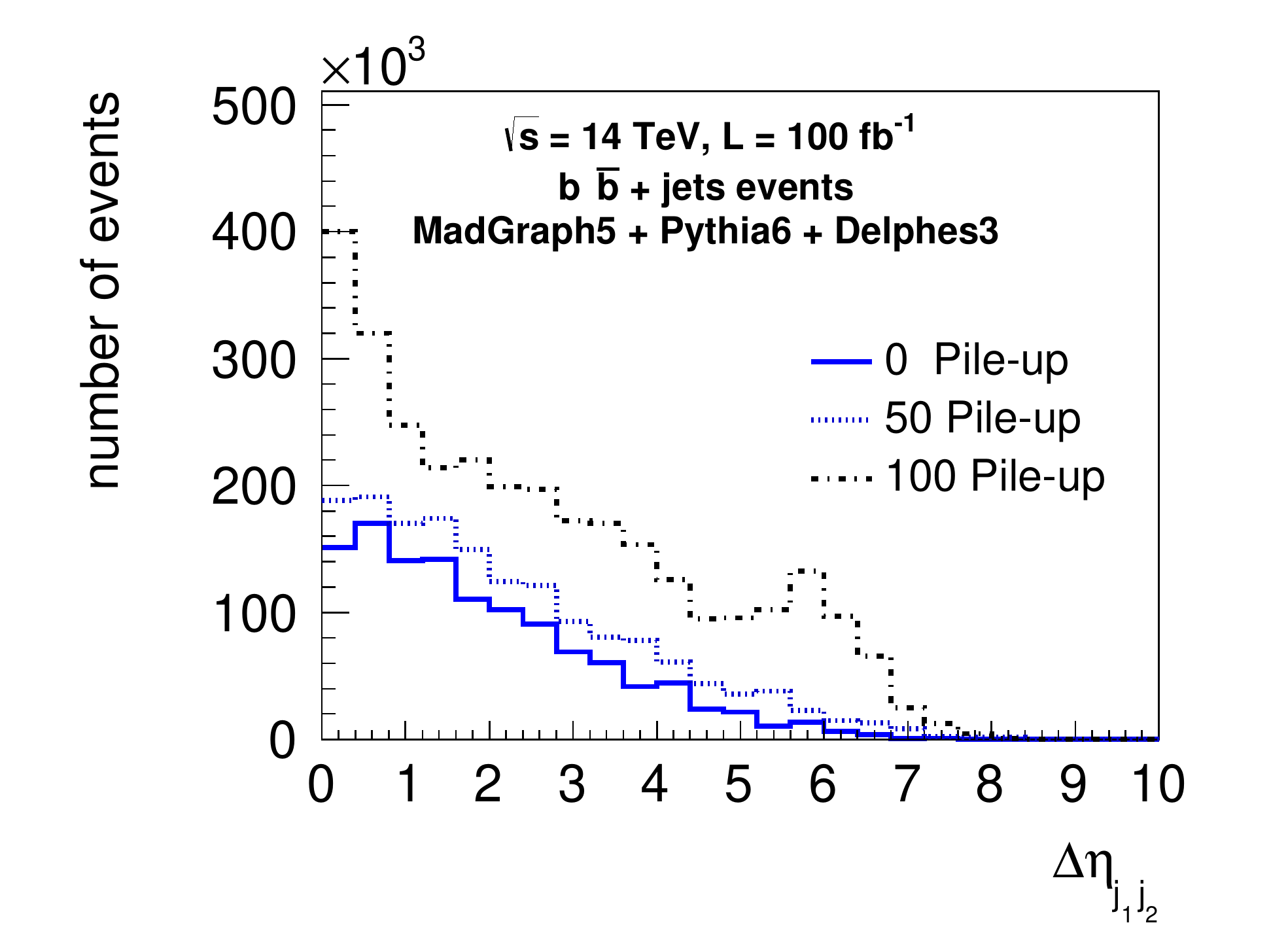}
\end{minipage}
\caption{Difference in pseudo-rapidity of the two most energetic reconstructed light jets for the signal (left) and the background (right) with 0, 50 and 100 average pile-up interactions. Histograms are normalized to the expected number of events predicted by \MG at $\sqrt{s}=14$ TeV for an integrated luminosity of $L=100$ $fb^{-1}$.}
 \label{fig:rapgap}
\end{figure}

In figure~\ref{fig:rapgap} the $\Delta \eta_{j_1j_2}$ distribution is shown for the signal (left) and background (right) for different pile-up scenarios. The normalization corresponds to the total number of events expected to pass selection (1) for an integrated luminosity $L=100$ $fb^{-1}$ at $\sqrt{s}=14$ TeV. As expected, with increasing pile-up, a significant number of additional jets emerges, despite the pile-up subtraction procedure, which leads to an increase in the amount of events passing selection (1). In the signal sample, pile-up jets are then more often wrongly selected as prompt signal jets, leading to a depletion of the rapidity gap. Pile-up also tends to inflate the total background contribution. This aspect is relevant in particular in the tail of the distribution, which corresponds to the signal region. A significant excess of background events is observed at 100 pile-up in the signal region (at $\Delta \eta_{j_1j_2}$ $\approx$ 6). This feature corresponds to the poor calorimeter resolution and a low granularity in the region $\eta > 2.5$, which leads to the appearance of several additional jets. 

\begin{figure}[tbp]
\centering
\begin{minipage}{.5\textwidth}
  \centering
  \includegraphics[width=1\linewidth]{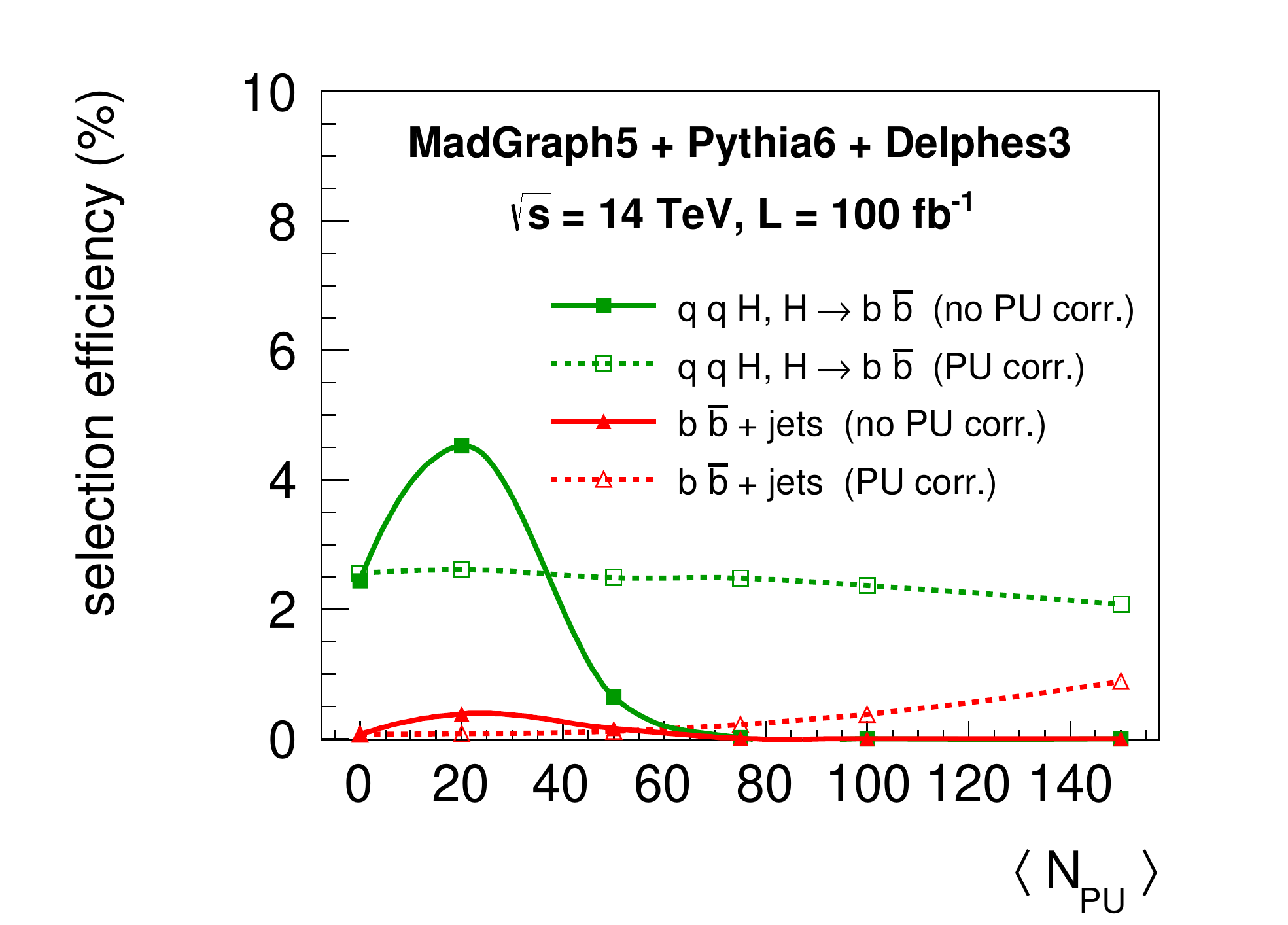}
\end{minipage}%
\begin{minipage}{.5\textwidth}
  \centering
  \includegraphics[width=1\linewidth]{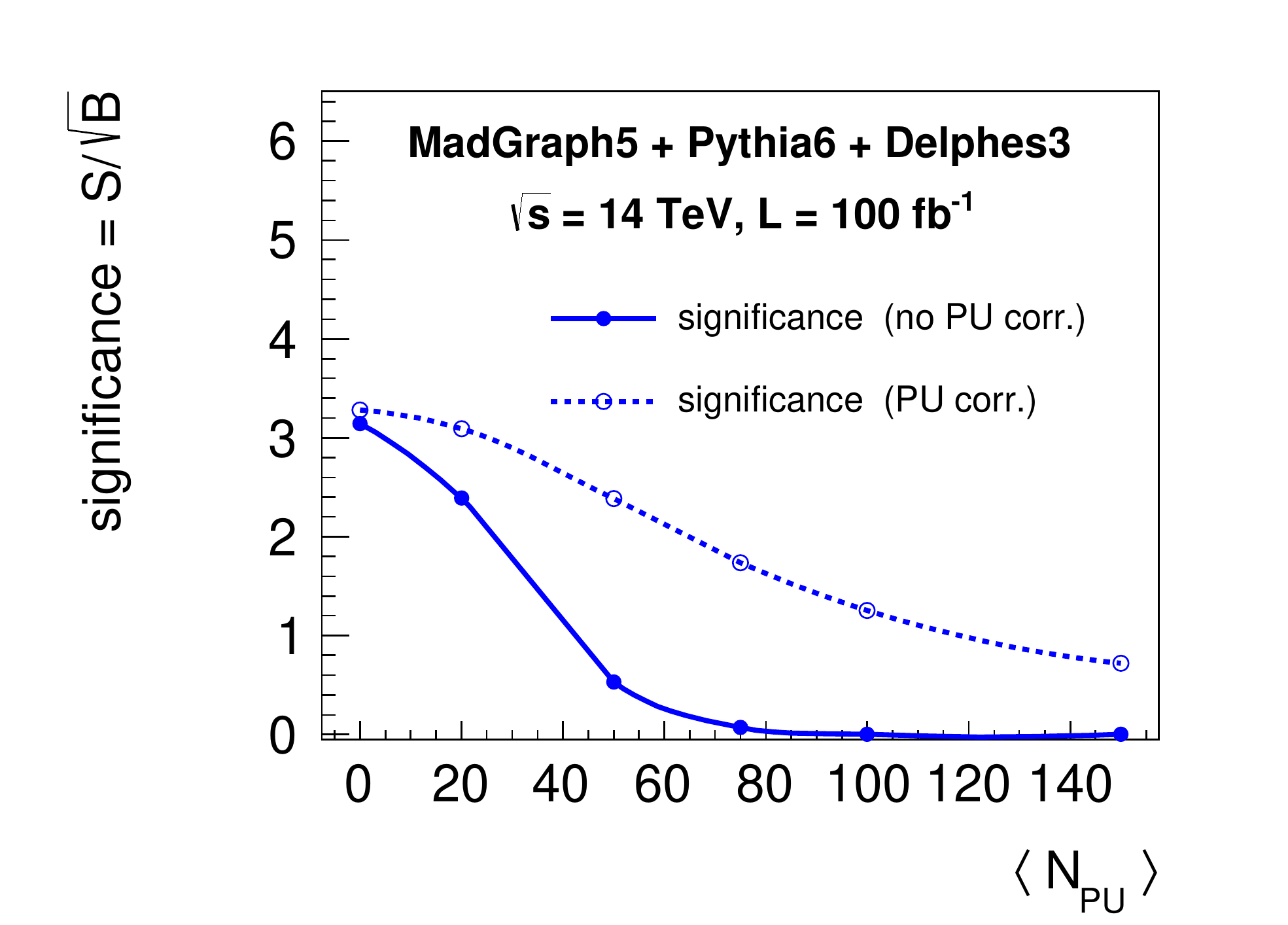}
\end{minipage}
\caption{Signal and Background total selection efficiencies (left) and Significance = $\frac{S}{\sqrt{B}}$ (right) as a function of the average number of pile-up interactions, with and without applying jet pile-up subtraction.}
 \label{fig:effsig}
\end{figure}

The total selection efficiency is shown in figure~\ref{fig:effsig} (left) for both signal and background. If jet pile-up subtraction is not applied, the efficiency rapidly grows as a function of pile-up until 20 pile-up interactions and decreases at higher pile-up. The increase is due to the emergence of additional jets, as explained earlier. However, when pile-up and the number of jets become too important, the probability of finding another jet in between the two hardest jets increases, hence drastically decreasing the efficiency of selection (2). It is clear from figure~\ref{fig:effsig} (left) that the pile-up subtraction procedure heavily slows down this effect, and results in a smoother, but yet still present, dependence on the number of pile-up interactions. The improvement brought by pile-up subtraction can also be seen on the signal significance in figure~\ref{fig:effsig} (right). 

With this example, we have shown that \DELPHES can be used to estimate the impact of pile-up on LHC studies. We emphasize that the predictions stated in this short study, should be understood as qualitative rather than quantitative, as indeed, Delphes has not yet been compared to full simulation studies at extreme pile-up conditions. Indeed, once done, \DELPHES predictions may become more quantitative in that domain too. However, by the time experiments reach such pile-up conditions, experimental collaborations may find ways to cope with pile-up that are not foreseen in \DELPHES.

On the other hand, it should be noted that no fully parametric study could eventually account for the effects that were illustrated here, unless a prior parameterization was obtained from a full simulation study. One should emphasize that simple analysis techniques are used in this study, so that the results are in no way representative of the ultimate potential of the LHC multipurpose detectors.

\section{Conclusion}\label{sec:ccl}

We discussed the version 3.0 of \DELPHES, a framework designed to perform a fast and realistic simulation of a general purpose collider experiment. The new modular design of \DELPHES was presented, and we described the principles used for modeling the detector response and parameterizing the event reconstruction. 

We showed that \DELPHES 3.0 is able to produce realistic observables and is fully validated. It can thus be used to perform quickly realistic physics studies without in-depth knowledge of the technicalities of real experiments. 

\clearpage

\appendix
\section{Software implementation}\label{sec:software}

The \DELPHES software is a modular framework written in C++ and is based on the \ROOT analysis framework~\cite{bib:root}. It is fully integrated within the \MGZ~\cite{bib:madg} suite. It makes use of other external libraries such as \FASTJET~\cite{bib:fastjet}, ExRootAnalysis~\cite{bib:exroot} and ProMC~\cite{bib:promc}.
In the following the code structure and technical performance is discussed.

\subsection{Code structure}\label{sec:codeStructure}

The \DELPHES framework can be subdivided in the following subsystems:

\begin{itemize}
  \item \emph{Memory manager} minimizes the amount of memory allocations. It allows the user to create, destroy and clear all data collections used by other services and modules. It also clears all data collections produced by other services and modules between events in the event loop. 
  \item \emph{Configuration manager} stores the parameters for all modules and provides access by name to these parameters.
  \item \emph{Data manager} provides access by name to all data collections created by other services and modules.
  \item \emph{Universal object} represents all physics objects (particles, tracks, calorimeter towers, jets) with possibility to add user defined information.
  \item \emph{Modules} consume and produce collections of universal objects.
	\item \emph{Readers} read data from different file formats.
\end{itemize}

The modular system allows the user to configure and schedule modules via a configuration file, add modules, change data flow, alter output information. One can for instance store collections of the same physical object obtained with different algorithms, such as leptons with different isolation criteria or jets with different b-tagging criteria. Modules communicate entirely via collections of universal objects ({\TObjArray} of {\Candidate} four-vector like objects).

\subsection{Data Flow}\label{sec:dataflow}

\begin{figure}[tbp]
\begin{center}
\includegraphics[width=0.75\textwidth]{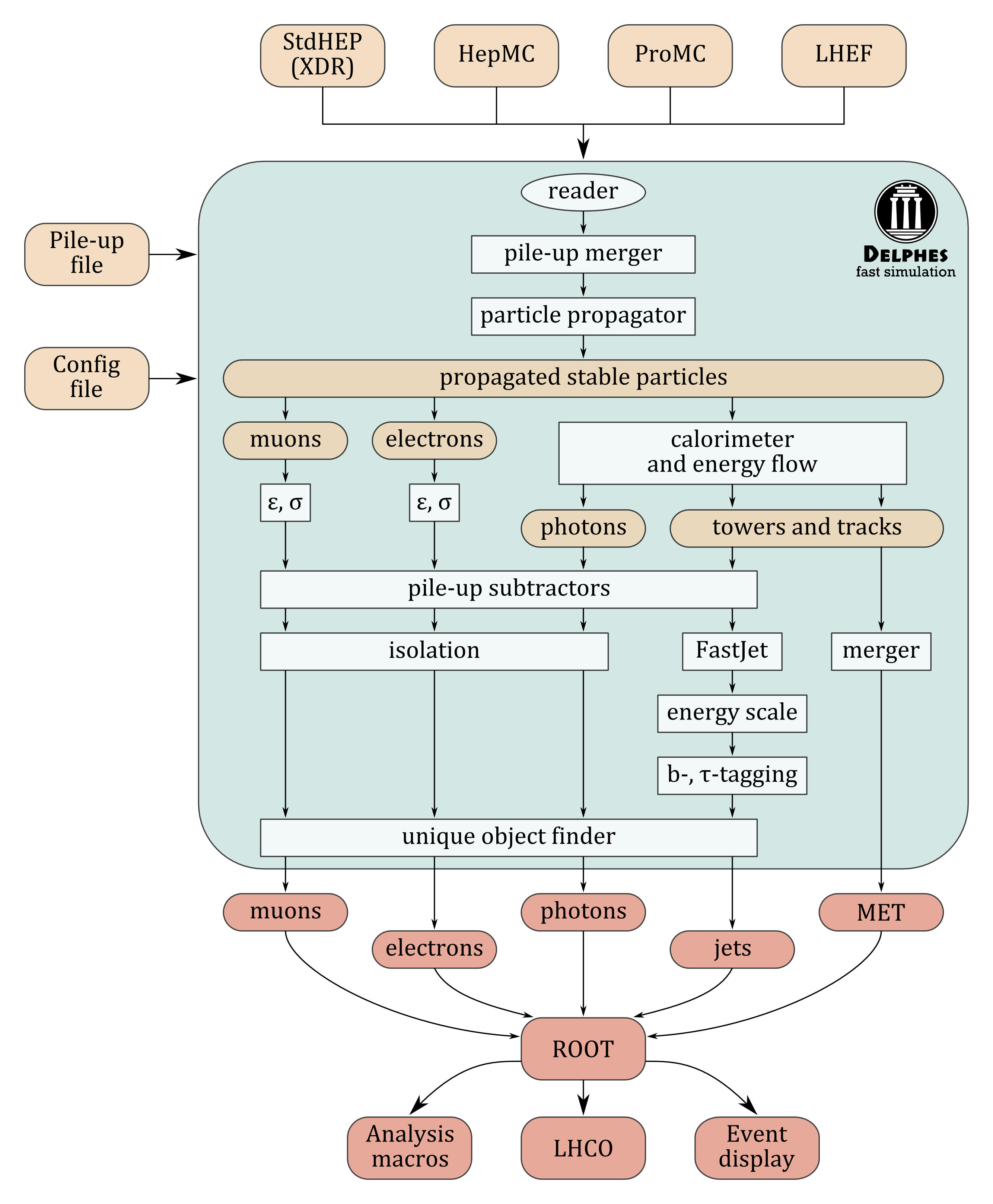}
\caption{Typical work-flow chart of the \DELPHES fast simulation. Event files coming from external Monte-Carlo generators are first processed by a reader stage (top). Pile-up events are then overlapped onto the hard scattering event. Long-lived particles are propagated to the calorimeters within
a uniform magnetic field. Particles reaching the calorimeters deposit their energy in the calorimeters. The particle-flow algorithm produces two collections of 4-vectors --- particle-flow tracks and towers. True photons and electrons with no reconstructed track that reach ECAL are reconstructed as photons. Electrons and muons are selected and their 4-vectors are smeared. Charged hadrons coming from pile-up vertices are discarded and the residual event pile-up density $\rho$ is calculated. The pile-up density $\rho$ is then used to perform pile-up subtraction on jets and on the isolation parameter for muons, electrons and photons. No pile-up subtraction is performed on the missing energy. At the final stage, the duplicates of the reconstructed objects are removed. The output data are stored in a \ROOT tree format and can be analyzed and visualized with the help of the \ROOT data analysis framework. The \ROOT tree
files can be also converted to the \LHCO file format. Each step is controlled by the configuration file.}
\label{fig:code_structure}
\end{center}
\end{figure}

A simplified data flow diagram is shown in figure~\ref{fig:code_structure}. The \DELPHES framework allows the access to data from different file formats
(ProMC~\cite{bib:promc}, HEPMC~\cite{bib:hepmc}, STDHEP~\cite{bib:stdhep} and the LesHouches event format
(LHEF)~\cite{bib:lhef}). Event files coming from external Monte-Carlo generators are first processed by a reader.
The Reader converts stable particles into a collection of universal objects. This collection is then processed by a series of modules
beginning with the pile-up merger module and ending with the unique object finder module. Finally, \DELPHES allows the user to store and analyze
events in a \ROOT tree format~\cite{bib:root}.

\ROOT tree objects are created from particles generated by a Monte-Carlo generator and from objects produced by \DELPHES
(physics objects like jets, electrons, muons, {\etc}). For uniformity, each branch is represented by a {\TClonesArray}.
If a branch contains a single entry per event (for example the $E_T^{miss}$), the branch is then represented by a {\TClonesArray}
with only one entry. Objects stored in the tree are linked by means of {\TRef} pointers or {\TRefArray} (array of pointers).
More documentation on the content of the \DELPHES output \ROOT tree is available on the \DELPHES website~\cite{bib:delphesweb}.

\begin{figure}[tbp]
\begin{center}
\includegraphics[width=1\linewidth]{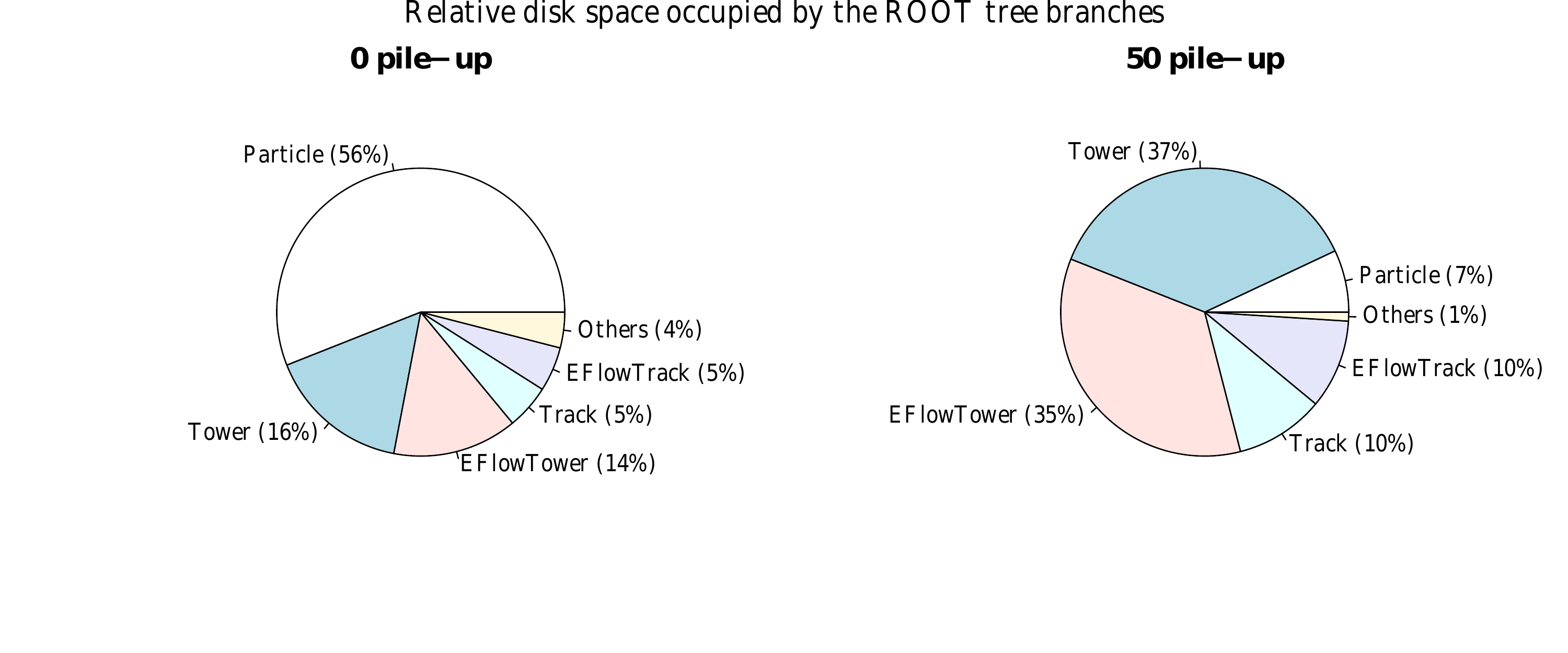}
\caption{Relative disk space occupied by the ROOT tree branches for a sample of $\rm{t\bar{t}}$+jets events without pile-up (left) and with 50 average pile-up interactions (right). As a reference, the typical disk space of $\rm{t\bar{t}}$+jets events reconstructed in \DELPHES is 30 kB/event without pile-up and 220 kB/event for events with 50 pile-up interactions.}
\label{fig:filesize}
\end{center}
\end{figure}

Relative disk space occupied by the ROOT tree branches for a sample of $\rm{t\bar{t}}$+jets events without pile-up and with 50 average
pile-up interactions is shown in figure~\ref{fig:filesize}. The Particle, Tower and EFlowTower branches occupy approximately 80\% of the total disk space.
If the available disk space is limited and if the information stored in these branches is not required for a particular analysis, the output file size can
be significantly reduced by disabling these branches in the configuration file.

\subsection{Technical Performance}\label{sec:performance}

The main motivation for a tool like \DELPHES is to minimize the resources needed on top of those used for event generation: small memory footprint,
efficient usage of CPU and small file size.

\begin{figure}[tbp]
\centering
\begin{minipage}{.5\textwidth}
  \centering
  \includegraphics[width=1\linewidth]{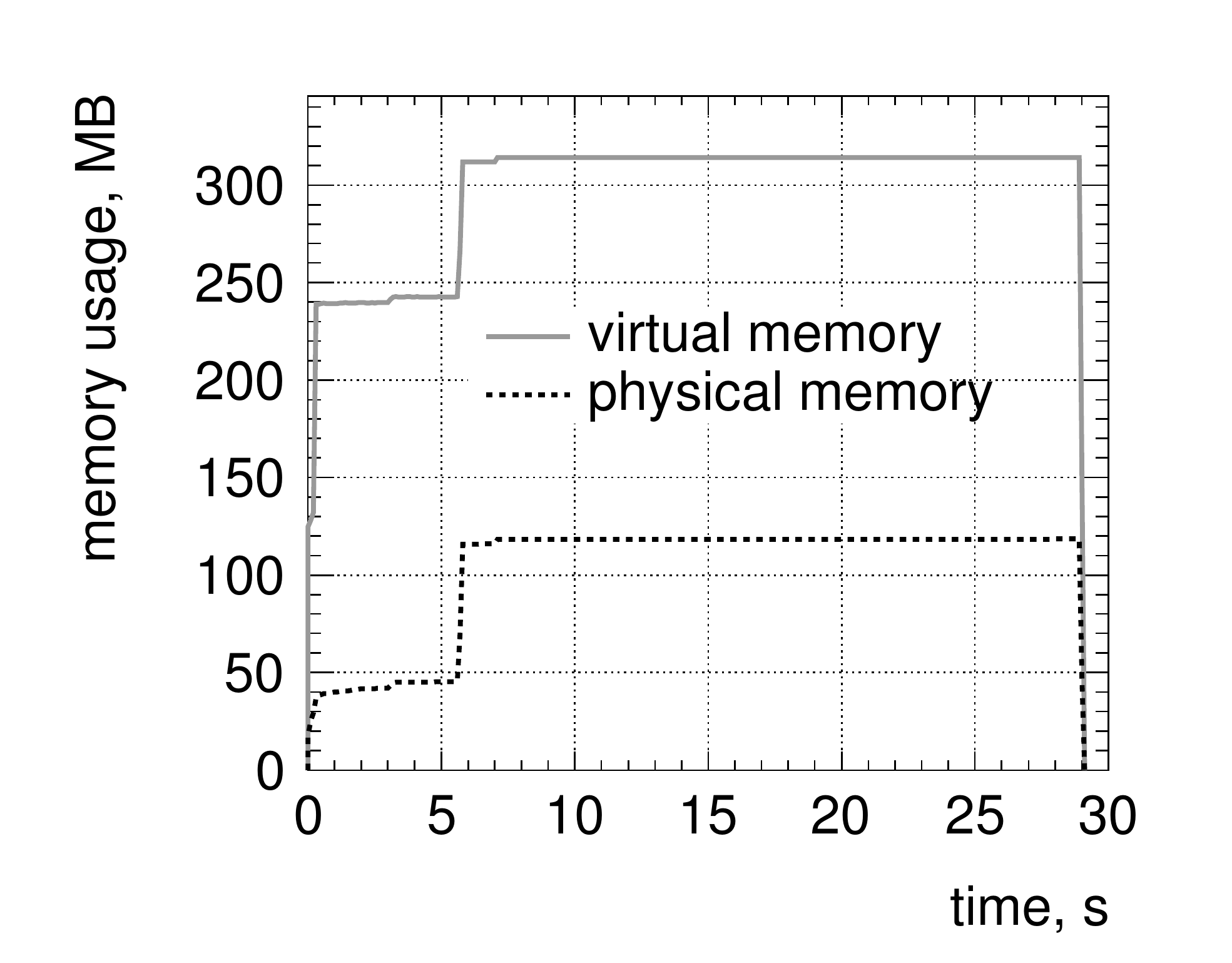}
\end{minipage}%
\begin{minipage}{.5\textwidth}
  \centering
  \includegraphics[width=1\linewidth]{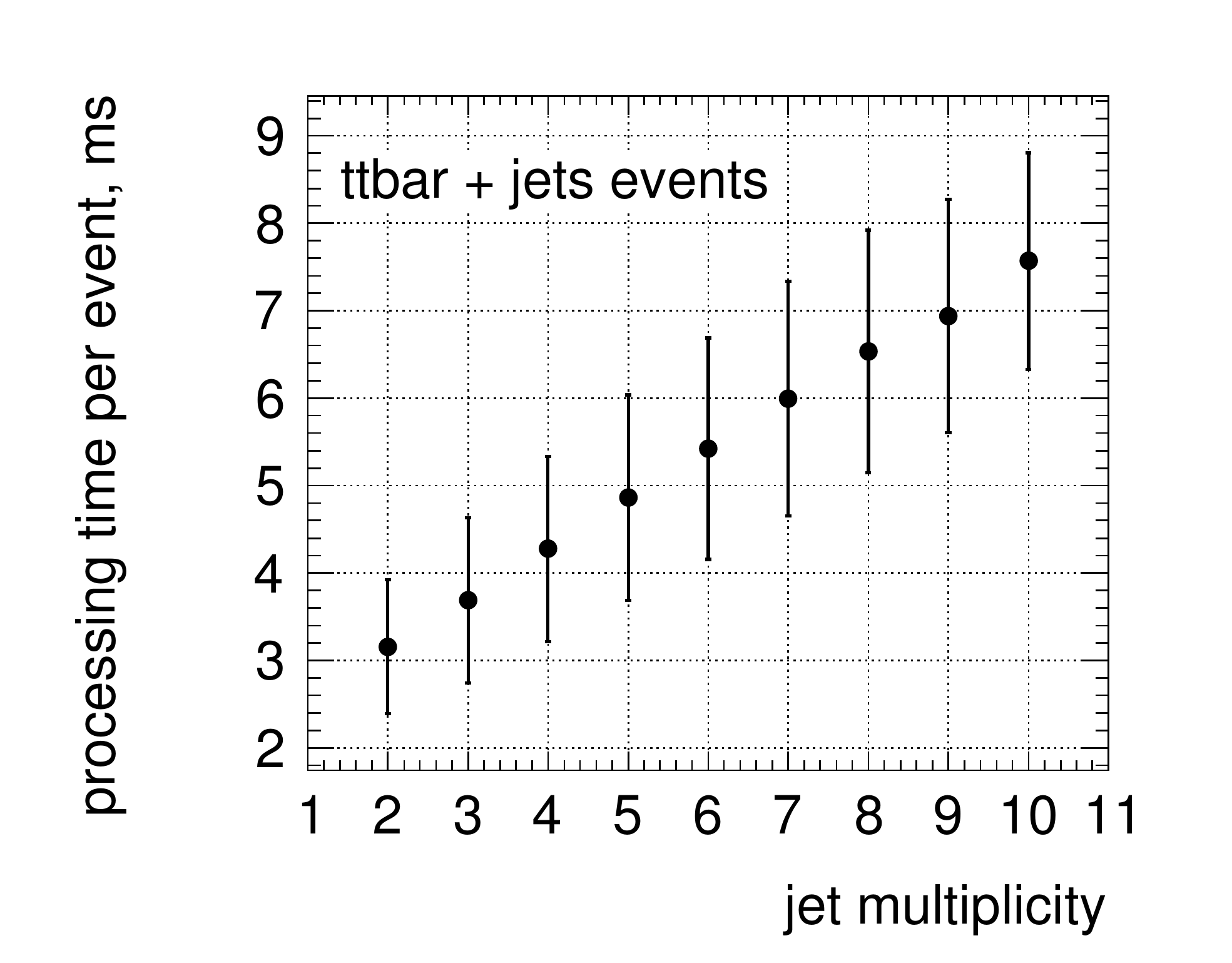}
\end{minipage}
\caption{Left: memory usage as function of time after the program start. Right: processing time per event as function of jet multiplicity on inclusive $\rm{t\bar{t}}$ events.
The vertical error bars represent RMS deviation of the time measurements in one bin.}
\label{fig:performance}
\end{figure}

Figure~\ref{fig:performance} illustrates how well \DELPHES achieves these goals. Memory usage does not exceed a few hundred megabytes
and remains constant after the initial memory allocation (figure~\ref{fig:performance}, left). The processing time as a function of the reconstructed jet
multiplicity is shown in figure~\ref{fig:performance}~(right) for $\rm{t\bar{t}}$+jets events. Processing time can be as low as a few milliseconds per event
and is expected to follow the scaling law of the underlying jet reconstruction algorithm.
For comparison, a typical $\rm{t\bar{t}}$ event takes approximately 80 seconds of CPU time for full simulation of the CMS detector and event reconstruction,
the fast simulation of the CMS detector takes approximately 1.6 seconds per event for the full chain~\cite{bib:cmsfastsim}.

\begin{figure}[tbp]
\begin{center}
\includegraphics[width=1\linewidth]{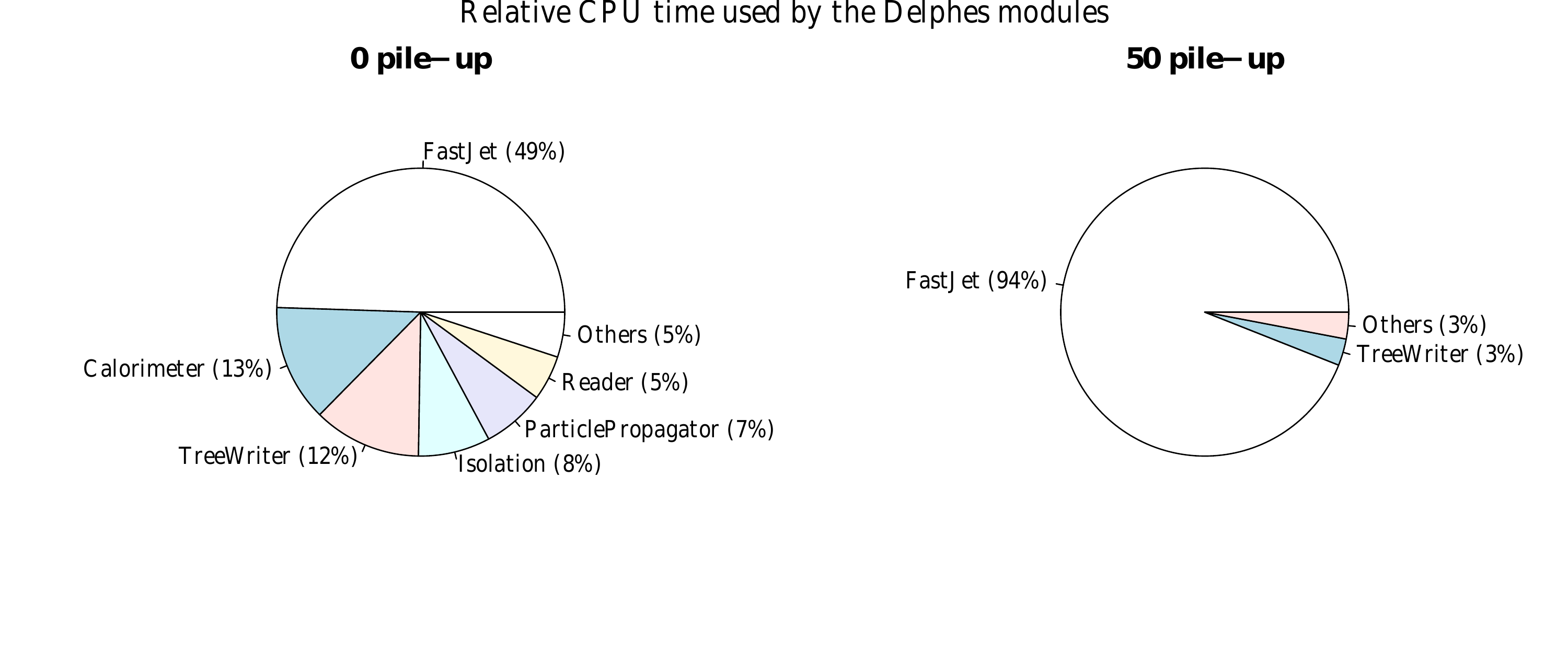}
\caption{Relative CPU time used by the Delphes modules while processing a sample of $\rm{t\bar{t}}$+jets events without pile-up (left) and with 50 average pile-up interactions (right).}
\label{fig:profile}
\end{center}
\end{figure}

Relative CPU time used by the \DELPHES modules while processing a sample of $\rm{t\bar{t}}$+jets events is shown in figure~\ref{fig:profile}.
Most of the CPU time is used by the jet reconstruction module (\FASTJET). This module uses approximately half of the total CPU time
for the events without pile-up and more than 90\% of the total CPU time for the events with 50 average pile-up interactions.
It should be noted that in the latter case \FASTJET is used for the jet reconstruction and for the residual event pile-up
density calculation. So, if there is any significant improvement potential, it lies in improving the performance of the
jet reconstruction and of the residual pile-up subtraction.

\acknowledgments

First, we would like to thank S\'everine Ovyn and Xavier Rouby for their invaluable contribution to the early \DELPHES development. We wish to thank the \DELPHES users for continuously providing feedback and valuable insights. We are very grateful to the \MG team for having integrated \DELPHES within \MG. Finally, we wish to thank the members of the CP3 for providing support and useful discussions. We acknowledge the support from the FNRS and the IAP Program, BELSPO VII/37. This work is partly supported by the IISN convention 4.4503.13.

\end{document}